
\font\tfo=cmbx9 scaled \magstep3
\input harvmac
\line{\hfill CERN-TH 6702/92}
\bigskip
\bigskip
\centerline{\tfo Conformal Symmetry and Universal Properties}
\medskip
\centerline{\tfo of Quantum Hall States }
\bigskip
\bigskip
\centerline{ Andrea CAPPELLI${}^a$\footnote{*}{\rm
      On leave of absence from I.N.F.N., Firenze, Italy.},
\ Gerald V. DUNNE${}^b$,}
\centerline{ Carlo A. TRUGENBERGER${}^a$ and Guillermo R. ZEMBA${}^a$}
\bigskip
\centerline{\it ${}^a$\ Theory Division\footnote{**}{\rm
      Bitnets CAPPELLI, CAT, ZEMBA at CERNVM.},
      C.E.R.N., 1211 Geneva 23, Switzerland}
\centerline{\it ${}^b$\ Dept. of Physics, University of Connecticut,
2152 Hillside Road, Storrs, CT 06268 USA}
\vskip .4in
\centerline{\bf ABSTRACT}
\vskip .2in
The low-lying excitations of a quantum Hall state on a disk geometry
are edge excitations. Their dynamics is governed by a conformal field
theory on the cylinder defined by the disk boundary and the time
variable.
We give a simple and detailed derivation of this conformal field theory
for integer filling, starting from the microscopic dynamics of
$(2+1)$-dimensional non-relativistic electrons in Landau levels.
This construction can be generalized to describe Laughlin's
fractional Hall states via chiral bosonization, thereby making
contact with the effective Chern-Simons theory approach.
The conformal field theory dictates the finite-size effects in the energy
spectrum. An experimental or numerical verification of these universal
effects would provide a further confirmation of Laughlin's theory
of incompressible quantum fluids.

\vskip .3in
\vfill
\noindent CERN-TH 6702/92

\Date{October 1992}

\newsec{Introduction}
Laughlin's picture
\ref\lor{R. B. Laughlin, {\it Phys. Rev. Lett.} {\bf 50} (1983) 1395.}
of {\it incompressible quantum fluids} is
the basis of our understanding of the plateaus in the quantum
Hall effect (QHE)
\ref\qhe{R. E. Prange and S. M. Girvin eds., {\it ``The Quantum
Hall effect''}, Springer, New York, (1990).}
\ref\wu{Y. S. Wu, {\it ``Topological Aspects of
the Quantum.Hall Effect''}, in {\it ``Physics, Geometry and
Topology''}, H. C. Lee ed., Proc. Banff Conf. 1989, Plenum Press,
New York (1990).}
\ref\frad{E. Fradkin, {\it ``Field Theories of Condensed Matter
systems''}, Addison-Wesley, New York, (1991).}.
Planar electrons in a uniform magnetic field $B$ behave as a
rigid droplet of liquid for specific rational values of the
{\it filling fraction} $\nu=2\pi\rho/B$ ($\rho=$density).
For these values, the electrons assume a
symmetric configuration with uniform density:
excitations about such a configuration face an energy gap
and hence the liquid is incompressible.

In systems with boundaries, as in the disk geometry we shall
mostly consider here, the only
possible low-energy excitations reside on the edge of the droplet.
The edge can change shape at constant density, and thus the area
of the droplet stays constant. This is the semiclassical picture for
the infinite symmetry of the incompressible ground
state under area-preserving diffeomorphisms
\ref\ctz{A. Cappelli, C. A. Trugenberger, and
G. R. Zemba, {\it ``Infinite Symmetry in the Quantum Hall Effect''},
CERN preprint 6516/92.}
{}\footnote{*}{See also refs.
\ref\wadia{S. Iso, D. Karabali and B. Sakita, {\it ``Fermions in the
Lowest Landau Level: Bosonization, $W_\infty$ Algebra, Droplets, Chiral
Boson''}, CCNY-HEP-92/6 preprint;
A. Dhar, G. Mandal and S. R. Wadia, {\it ``Classical Fermi Fluid and
Geometrical Action for c=1''}, IASSNS-HEP-91/89 preprint.}
in connection with string theory.}.

Starting with Halperin
\ref\halpe{B. I. Halperin, {\it Phys. Rev.} {\bf B25} (1982) 2185.},
a number of authors
have been studying the dynamics of edge excitations
\ref\wen{X. G. Wen, {\it ``Gapless Boundary Excitations in the Quantum
Hall States and the Chiral Spin States''}, preprint NSF-ITP-89-157,
{\it Phys. Rev. Lett.} {\bf 64} (1990) 2206;
B. Block and X. G. Wen, {\it Phys. Rev.} {\bf B42} (1990) 8133;
X. G. Wen, {\it Phys. Rev.} {\bf B41} (1990) 12838,
D. H. Lee and W. G. Wen, {\it Phis. Rev. Lett.} {\bf 66} (1991) 1765.}
\ref\stone{M. Stone, {\it Ann. Phys.} (NY) {\bf 207} (1991) 38,
{\it Phys. Rev.} {\bf B42} (1990) 8399,
{\it Int. Jour. Mod. Phys.} {\bf B 5} (1991) 509.}
\ref\fro{J. Fr\"ohlich and A. Zee, {\it Nucl. Phys.} {\bf B364}
(1991) 517; J. Fr\"ohlich and T. Kerler, {\it Nucl. Phys.}
{\bf B 354} (1991) 369.}
\ref\wenrev{For a review see: X. G. Wen, {\it Int. Jour. Mod. Phys. }
{\bf B6} (1992) 1711.}.
These live in one space dimension, the boundary of the
circular droplet, and they are chiral due to the external
magnetic field. For realistic samples, the radius $R$ of the
circle is much larger than the typical microscopic scale
given by the cyclotron radius $\ell = \sqrt{2\hbar c /eB}$.
Therefore, one can study the long-range effective theory of the
edge excitations. This theory possesses universal properties which
can be observed experimentally, as in the case of effective theories
of critical phenomena.

In this theory, we are expecting a large symmetry
connected to the already mentioned infinite symmetry of the
ground state. Actually, Laughlin's theory
\ref\laugha{R. B. Laughlin, {\it ``Elementary Theory: the
Incompressible Quantum Fluid''}, in \qhe .}
exhibits many indications of long-range coherence, which
can be accounted for by the presence of a high degree of
symmetry. The excitations are {\it anyons}, particles
with fractional statistics
\ref\wil{For a review see: F. Wilczek, {\it ``Fractional
Statistics and Anyon
Superconductivity''}, World Scientific, Singapore, (1990).}.
Their wave-functions acquire non-trivial phases under mutual exchange,
no matter what the separation distance
\ref\laughb{R. B. Laughlin, ``Fractional Statistics in the Quantum
Hall effect'', in \wil .}.
This is indeed a consequence of infinite {\it conformal invariance}
\ref\bpz{A. A. Belavin, A. M. Polyakov and A. B. Zamolodchikov,
{\it Nucl. Phys.} {\bf B 241} (1984) 333.}
of the chiral $(1+1)$-dimensional theory
of the edge excitations.

Conformal invariance implies {\it algebraic long-range order}
(power-law behaviour) in correlators of excitations localized
around the edge.
Precisely as in the low-temperature phase of
the $XY$ model, the long-range order should be understood in
terms of conformal invariance and not of spontaneous symmetry
breaking
\ref\xy{P. di Francesco, H. Saleur and J.-B. Zuber, {\it J. Stat. Phys.}
{\bf 49} (1987) 57.}.
We believe that this point of view on long-range order is more
natural than the one proposed in refs.
\ref\odlro{S. M. Girvin and A. H. Mac Donald, {\it Phys. Rev. Lett.}
{\bf 58} (1987) 1252;
N. Read, {\it Phys. Rev. Lett.} {\bf 62} (1989) 86.}.

Previous studies of conformal invariance and edge excitations in the
QHE were based mainly on the relation between Chern-Simons gauge
theories and conformal field theories (CFT) \wen \fro.
The bulk dynamics of an incompressible fluid can be described by an
effective long-range field theory of an Abelian Chern-Simons gauge
field. This field is the dual vector of the matter Hall current,
while the Hall conductivity gives the value of the Chern-Simons
coupling constant. On a disk geometry, the boundary gauge degree of
freedom (a chiral bosonic field) becomes physical and describes the
edge excitations in a bosonic language.

While this approach leads to an elegant formulation of the edge
dynamics, the relation between the bosonic edge theory and the original
problem of $(2+1)$-dimensional non-relativistic electrons in
Landau levels remains to be clarified.
The key idea to this end was provided by Stone, who understood that the
$(2+1)$-dimensional fermion goes into a $(1+1)$-dimensional
relativistic chiral charged fermion (Weyl fermion), whose
Dirac sea corresponds to the filled Landau level \stone.
Here we refine Stone's idea by introducing a confining pressure
via a novel procedure. For filling $\nu=1$, this allows a
straightforward derivation of the fermionic boundary theory
and its conformal symmetry directly from the microscopic physics
of the Landau levels.
We pay special attention to the details of the limiting procedure,
including the role of boundary conditions.
The resulting edge dynamics is a $c=1$ CFT describing a
Weyl fermion with Neveu-Schwarz boundary conditions
\ref\gins{For a review see: P. Ginsparg, {\it ``Applied Conformal
Field Theory''}, in {\it ``Fields, Strings and Critical Phenomena''},
Les Houches 88, E. Brezin and J. Zinn-Justin eds., North-Holland,
Amsterdam (1990).}.
We also show how to deform the CFT in order to include the gaps
for charged excitations on a disk geometry.
Bosonization leads to the Floreanini-Jackiw model
\ref\flore{R. Floreanini and R. Jackiw, {\it Phys. Rev. Lett.}
{\bf 59} (1987) 1873.}
of a chiral boson, which can be extended by introducing an additional
coupling constant. By this procedure we make contact with the
Chern-Simons approach.
The simplicity of our limiting procedure leads to a clear picture
of the interplay between $(2+1)$- and $(1+1)$-dimensional physics.

While confirming the results of previous investigations
\wen\stone\fro\wenrev\ about
the charge and statistics of excitations, we stress the role of
{\it finite-size effects} in the physics of the QHE.
Indeed, conformal invariance on the cylinder (the spatial boundary circle
times the time coordinate) implies a non-trivial result
\ref\cardy{J. L. Cardy, {\it ``Conformal Invariance and Statistical
Mechanics''}, in {\it ``Fields, Strings and Critical Phenomena''},
Les Houches 88, E. Brezin and J. Zinn-Justin eds., North-Holland,
Amsterdam (1990).}:
the $O(1/R)$-terms ($R$ being the radius of the droplet)
in the energy spectrum are {\it universal}.
These include both the {\it Casimir ground-state energy} and
the $1/R$-corrections to the gaps of excitations.
The former represents a quantum pressure of the electron
liquid and is determined by the central charge $c$ of the CFT.
The latter are determined by the conformal dimensions of the primary
fields.

The results of conformal invariance for charge and statistics
of excitations are in total agreement with Laughlin's theory and
can be considered an a posteriori justification of this theory
based on symmetry principles.
The $1/R$-terms in the energy spectrum are related to
the charge and statistics of the excitations.
Therefore, their comparison with experiments
and numerical simulations would provide a direct confirmation
of Laughlin's theory.
Moreover, the $1/R$-spectroscopy of hierarchical states is a
promising tool for uncovering their dynamics.

Our straightforward relation between boundary and bulk theories
also helps understanding some works which developed a formal analogy
between conformal correlators and the Laughlin wave function
\ref\fub{S. Fubini, {\it Mod. Phys. Lett.} {\bf A6} (1991) 347;
S. Fubini and C. A. L\"utken, {\it Mod. Phys. Lett.} {\bf A 6} (1991) 487.}
\ref\dlt{G. V. Dunne, A. Lerda and C. A. Trugenberger,
{\it Mod. Phys. Lett. } {\bf A 6} (1991) 2819.}
\ref\napo{C. Cristofano, G. Maiella, R. Musto and F. Nicodemi,
{\it Phys. Lett.} {\bf B 262} (1991) 88, {\it Mod. Phys. Lett.} {\bf
A 6} (1991) 1779, {\it Mod. Phys. Lett.} {\bf A 6} (1991) 2985,
{\it ``The Quantum Hall Effect at Arbitrary Rational Filling: a proposal''},
Napoli preprint DSF-T-92/05.}
\ref\moore{G. Moore and N. Read, {\it Nucl. Phys.} {\bf B 360} (1991) 362,
{\it ``Non-abelions in the Fractional Quantum Hall Effect''},
proc. $4$-th Yukawa Int. Seminar, Kyoto, July 1991,
{\it Prog. Theor. Phys. Supp.} {\bf 107} (1992), Y. Nagaoka ed..}.
In short, the conformal correlators are defined on the boundary
circle, but also determine the analytic part of correlators in the plane
by analytic continuation. Therefore, most of the results obtained
from the formal analogy are actually correct.
The {\it ``topological order''}
\ref\topo{X. G. Wen, {\it Phys. Rev.} {\bf B40} (1989) 7387,
{\it Int. Jour. Mod. Phys.} {\bf B2} (1990) 239;
X. G. Wen and Q. Niu, {\it Phys. Rev.} {\bf B41} (1990) 9377.}
of the Laughlin wave function on a
toroidal geometry is due to a well-known interplay between conformal
and modular invariance on the torus \cardy~,
which has been discussed specifically in refs. \napo
\ref\iengo{R. Iengo and K. Lechner, {\it Phys. Rep.} {\bf C 213}
(1992) 179.}.

The outline of the paper is as follows. In section $2$, we review
the Landau level problem in the case of filling fraction $\nu = 1$,
and modify the Hamiltonian so as to include the boundary effects in
a realistic finite sample. In section $3$, we discuss the
thermodynamic limit $R\to\infty$, and derive the $(1+1)$-dimensional
theory of the Weyl fermion and its conformal invariance. In
section $4$, we discuss the physical consequences of this theory
and the universal finite-size effects of the quantum
Hall system. In section $5$, the bosonization
of the Weyl fermion is carried out, by using the canonical theory of
the chiral boson. In section $6$, this theory is used to describe
the edge excitations of the simplest fractional Hall states.
A more technical result is  presented in Appendix A. This reports
a direct derivation of long-range order in the density matrix on the
boundary.
\bigskip
\vfill\eject
\newsec{Integer Quantum Hall state for a finite sample}
\bigskip
\noindent{\bf 2.1 Landau levels}
\bigskip
We consider spin-polarized, planar electrons of mass $m$ and electric
charge $e$ in an external, uniform, magnetic field $B >0$ (units
$\hbar =1, c=1$). The one-particle Hamiltonian is given by
\eqn\elec{
H=-{1\over{2m}}
\left( {\bf \nabla}
- i e {\bf A} \right )^{2} \ - {e B \over{2m}}\ ,}
where the second term represents the Pauli interaction. Since we
shall consider circular geometries, we choose to use the
the symmetric gauge
$ A_i ={B\over 2}\varepsilon^{ij} x^{j}, \ \ i,j=1,2 \ ,$
for the external vector potential.
The fundamental scale set by the external magnetic field is the
magnetic length, $\ell=\sqrt{2/eB}$.
In the following, we will make extensive use of the complex notation
$z=x^{1}+ix^{2}$, $\bar z=x^{1}-ix^{2}$, $\partial =\partial
/\partial z$, $\bar \partial =\partial /\partial \bar z$.
By introducing two commuting sets of harmonic oscillator operators
\eqn\eab{\eqalign{
d &={z\over 2\ell} +\ell {\bar \partial}  \ ,
\qquad d^{\dag } ={{\bar z}\over 2\ell} - \ell \partial  \ ,
\qquad [d,d^{\dag} ]=1 \ , \cr
c &={\bar z\over 2\ell} +\ell \partial  \ ,
\qquad c^{\dag } ={{ z}\over 2\ell} - \ell {\bar\partial}  \ ,
\qquad [c,c^{\dag} ]=1 \ , \cr}}
the Hamiltonian \elec\ and the canonical
angular momentum $J= -i x^{i} \varepsilon^{ij} \partial_{j}$ can
be rewritten as
\eqn\llh{\eqalign{H &= \ \omega\ d^{\dag }d\ , \cr
J &= \ c^{\dag }c -d^{\dag }d \ ,\cr }}
where $\omega=eB/m$ is the cyclotron frequency. Since the operators
$c$ and $d$ commute, the spectrum consists of infinitely degenerate
levels of energy $\varepsilon_{n}=\omega n$: these are called the Landau
levels. The degenerate states in one Landau level are
characterized by the angular momentum eigenvalue $l$.
Wave functions $\psi$ of the first Landau level
satisfy $d \psi=0$ and have vanishing energy due to the additional
Pauli interaction in \elec. A complete basis is given by
\eqn\psd{\psi_{l}({z,{\bar z}})=
{1\over{\ell\sqrt{\pi}}}\
{1\over{\sqrt{l!}}}\ \left({z\over\ell}\right)^{l}\
{\rm e}^{\scriptstyle - |z|^2 /2\ell^2} .}
Note that the radial part of these wavefunctions is sharply peaked around a
radius $r_{l}=\ell\sqrt{l}$.
Higher Landau levels basis states can be
obtained by repeated action of the operators $d^{\dag}$ and
$c^{\dag}$ on $\psi_0$:
\eqn\psdhll{\eqalign{ \psi_{n,l}({z,{\bar z}})\ &=\
{{(c^{\dag})^{l+n}}\over{\sqrt{(l+n)!}}}
{{(d^{\dag})^{n}}\over{\sqrt{n !}}}
\psi_{0} (z,{\bar z}) \cr
&={1\over\ell}\ {\sqrt{n!\over{\pi (l+n)!}}}
\left( z\over\ell \right)^l \
L^{l}_{n}\left( {|z|^{2}\over\ell^2} \right) \
{\rm e}^{\scriptstyle  -|z|^2/2\ell^2 } . \cr}}
Here $n=0,1,\dots$ labels the energy level, $l+n \ge 0$, and
$L^{l}_{n}(x)$ are the generalized Laguerre polynomials
\ref\tables{I. S. Gradshteyn and I. M. Ryzhik, {\it ``Table of Integrals,
Series and Products''}, Academic Press, New York (1980).}.

The many-body problem of $N$ electrons is described,
in second quantization, by the Hamiltonian
\eqn\handn{H= {1\over{2m}} \int d^{2} {\bf x} \
{\left ( D_{i} \Psi \right )}^{\dag} {\left ( D_{i}
\Psi \right )}
-{1\over{2m}}\int d^{2} {\bf x}\ eB \rho \,}
with $D_{i}= \partial_{i} +i e A_{i}$ the covariant derivative
and $\rho({\bf x},t)= {\Psi}^{\dag}({\bf x},t) \Psi ({\bf x},t)$
the particle-number density.
The field operator $\Psi({\bf x},t)$ possesses an expansion in terms of
the single-particle energy and angular momentum eigenstates
$\psi_{n,l}({\bf x})$
\eqn\fielde{\Psi({\bf x},t) = \sum_{n=0}^{\infty}
\sum_{l=- n}^{\infty} a_{l}^{(n)}
\psi_{n,l}({\bf x})\ e^{-in\omega t}\ .}
The coefficients are fermionic
Fock annihilators and creators satisfying
\eqn\antic{\{ a_{k}^{(n)}, a^{(m)\dag}_{l} \}= \delta_{n,m}
\delta_{k,l} \ .}
with all other anticommutators vanishing. In the following we will
consider mostly the physics of the first Landau level, for which
the field operator takes the (time-independent) form
\eqn\fllfo{\Psi ({\bf x})\ = \sum_{l=0}^{\infty}\ a_{l}
\ \psi_{l}({\bf x}) \,}
with $\psi_{l}({\bf x})$ given by \psd\ and $a_l \equiv a^{(0)}_l$.
The property $d\psi_l =0$ can be also written as a
self-dual condition\footnote{*}{The corresponding
condition $D_{-}\Psi =0$ would hold for an external magnetic field with
the opposite sign.} on the field operator
\eqn\sdc{D_{+} \Psi ({\bf x})\ \equiv \left( D_{1} + i D_{2} \right)
\Psi ({\bf x})\ = 0\ .}
\bigskip
\bigskip
\noindent
{\bf 2.2 Incompressible ground state}
\bigskip
In the following we are interested in universal, long-range
properties of the many-body state consisting of one completely
filled Landau level on a disk geometry. By this we mean a state
$|\Omega\rangle$
in which all angular momentum states of the first Landau level
are occupied up to (and including) a maximal angular momentum $L$
(the total number of particles is then $N=L+1$):
\eqn\defome{|\Omega\rangle\ =\ a^{\dag}_0 a^{\dag}_1 \dots
a^{\dag}_L |0\rangle\ ,}
where $|0\rangle$ is the Fock vacuum.
Since the
single-particle angular momentum states are peaked around radii
$r_{l}=\ell\sqrt l$, the state under consideration consists of
a circular droplet of radius approximately given by
$R\simeq \ell\sqrt{L} \ $,
as is rendered pictorially in Fig. 1. This
configuration is clearly {\it incompressible}: a compression
of the droplet would lower its total angular momentum and face
an energy gap $\omega$, since at least one electron would be
promoted to the next Landau level. The incompressibility is
reflected in the spatial distribution of the matter.
The expectation value of the density is readily computed in the
second-quantized formalism,
\eqn\exprho{\langle \Omega|\rho({\bf x})|\Omega\rangle  =
{1\over \ell^2 \pi}\ {\rm e}^{-r^2/\ell^2 }\ \sum_{l=0}^{L}
{1\over l!} \left(  {r\over\ell} \right)^{2l} \,}
and is shown in Fig. 2. It is constant  for $r \ll  \ell\sqrt{L}$,
and drops rapidly to zero around $r \simeq \ell\sqrt L$.
Thus incompressibility implies uniform density.
Another quantity of interest is the expectation value of the
current
\eqn\curre{J^{i}({\bf x})\ = {1\over{2im}} \left\{{\Psi}^{\dag}({\bf x})
D_{i} \Psi ({\bf x})
-(D_{i} \Psi ({\bf x}))^{\dag} \Psi ({\bf x}) \right\} \ .}
This is easily obtained by combining \curre~ with
$\partial_{i} \rho={\Psi}^{\dag}D_{i}\Psi+(D_{i}\Psi)^{\dag}\Psi$
and using the self-dual condition \sdc. This gives
\eqn\expj{\langle\Omega|J^{i}({\bf x})|\Omega\rangle  = -
{1\over{2m}} \varepsilon^{ij}\partial_{j} \langle \Omega|\rho({\bf
x})|\Omega\rangle \,}
which is shown in Fig. 3. The current vanishes in the interior,
where the density is constant; however, the state supports a
transverse, circular current around the boundary of the droplet.
This is the edge current originally discussed by Halperin \halpe\
and
gives a first indication that non-trivial dynamics is concentrated
around the boundary. This boundary dynamics will be studied in
detail in the following.

We must distinguish between bulk and boundary interactions.
Here we are not interested in the bulk dynamics which creates the
incompressible ground state.
To understand this dynamics, one should consider the microscopic
interaction of electrons with impurities and their Coulomb interaction,
and then solve the strongly interacting problem.
Rather, we are going to assume that the incompressible ground state
is indeed formed. This corresponds to an equilibrium configuration
for the bulk interactions with a gap for density fluctuations.
Therefore, it is reasonable to neglect these in the study of
long-wavelength edge fluctuations.

The boundary interaction is described by a finite-size modification
of the one-body Hamiltonian of Landau levels.
This modification can be obtained by quantization in a finite disk
(or annulus) geometry  (Dirichlet boundary conditions) \halpe.
However, physically, it originates from the external potential due to the
neutralizing background which keeps the electrons together \laugha.
Both effects lift the degeneracy of the Landau levels near the boundary.
The angular momentum eigenstates in the first Landau level acquire a
spectrum $\varepsilon_{l}$ with $\Delta\varepsilon_l / \Delta l  > 0$ for
$l \sim L$. As a consequence, also decompressions, {\it i.e.},
transitions to states with higher angular momentum, cost energy and
$|\Omega \rangle$  is the unique ground state. Typically, the cost
in energy for transitions of a few electrons from states of angular
momentum $L - \delta l$ to states of angular momentum $L + \delta l$,
with $\delta l \ll L$, is much smaller than the cyclotron energy.
These {\it edge excitations} constitute the generic low-lying
excitations about the incompressible state $|\Omega\rangle$.

For $L \gg 1$ the spectrum $\varepsilon_{l}$ can be taken as approximately
linear in a region around $L$:
\eqn\linear{\varepsilon_l = \alpha  (l-L) + \beta \ .}
Since this spectrum originates dynamically from the neutralizing
background, the coefficients $\alpha$ and $\beta$ depend generically
on the particle number $N=L+1$.
In particular, one can consider their expansions in powers of
$\sqrt{L} \simeq R/\ell$, where $R$ is the size of the system.
As we now show, the dynamics of edge excitations is governed by a
{\it conformal field theory} if the region  of approximately linear
spectrum increases as $L \to \infty$ and if $\alpha$ possesses
an expansion
\eqn\alphaexp{\alpha= {v\over \ell\sqrt{L}} +
O\left( {1\over L} \right) \ \ .}
Here $v$ is the ``velocity of light'' of the emerging
relativistic theory with spectrum \hbox{$\varepsilon\propto vp$} with
$p=l/\ell\sqrt{L}$ the one-dimensional momentum.
Taking into account the general expansion of $\beta$,
\eqn\betaexp{\beta =\beta_0 -{v\mu\over \ell\sqrt{L} }
+ O\left( {1\over L} \right) \ \ ,}
the spectrum $\varepsilon_l$ up to order $O(1/\sqrt{L})$ takes the form
\eqn\betazero{\varepsilon_l\ =\ \beta_0 +
{v\over \ell\sqrt{L} }\left( l-L-\mu \right) }
with $\mu$ a dimensionless parameter playing the role of a
chemical potential.

At this point we must distinguish between a geometry with two
boundaries, i.e. an annulus, and the disk geometry with only one
boundary.
On an annulus geometry, the parameter $\beta_0$ becomes irrelevant
for all edge excitations since the spectrum $\varepsilon_l$
takes the generic form shown in Fig. 4.
Indeed, by considering $L_1,L_2 \gg 1$, $L_2/ L_1=\lambda=O(1)$,
stability requires that the Fermi levels at the two edges can differ
at most by terms of $O(1/\ell\sqrt{L_2})$.
Therefore, one can reabsorbe $\beta_0$ by an overall shift in the
zero of energy, thereby obtaining the spectra
\eqn\annulspec{ \varepsilon_l \ = \ \left\{
\eqalign{- {v_1\over \ell\sqrt{L_1} } \left( l-L_1 +\mu_1 \right) \ ,
& \qquad l\simeq L_1, \cr
{v_2\over \ell\sqrt{L_2} } \left(l-L_2 -\mu_2 \right) \ ,
& \qquad l\simeq L_2 \ .\cr} \right. }
At each edge there are two types of excitations: neutral particle-hole
excitations and charged excitations which amount to a transfer of
particles from one edge to the other.
Both are gapless since charge transfer costs only $O(1/\ell\sqrt{L_2})$
energy.
As we now show, in the thermodynamic limit $L_1,L_2 \to \infty$ all
these excitations span the Hilbert space of a $c=1$ CFT.
This is also the picture that emerges from the effective Chern-Simons
gauge theory approach \wen\fro.

The situation is different on a disk geometry.
Here the generic spectrum $\varepsilon_l$ looks as in Fig. 5 and
it costs a finite amount of energy to produce charged edge excitations,
since one has to transfer charge between the edge and the center of
the disk.
A positive charge on the edge corresponds to a quasi-hole in the
center of the disk and costs an energy $\beta_0 +O(1/\ell\sqrt{L})$.
A negative charge on the edge corresponds to a quasi-particle at
the center of the disk and costs an energy
$\Delta-\beta_0 +O(1/\ell\sqrt{L})$, where $\Delta$ is the gap
to the next energy band (for $\nu=1$ we have e.g. $\Delta=\omega$).
The gapfulness of the charged excitations spoils conformal invariance.
However, this is not a fatal problem; as we will show, the $c=1$ CFT
is still relevant insofar as it predicts the universal finite-size
corrections to the gaps and the static properties of correlators
on the edge.
\bigskip
\vfill\eject
\noindent
{\bf 2.3 Confining pressure via a boundary Hamiltonian}
\bigskip
In order to discuss quantitatively the effects described above,
we introduce a confining pressure by a novel procedure, which
also leads automatically to the mapping on a relativistic
CFT. We pick a circle of radius $R$ in the infinite plane
and we define a new Hamiltonian $H_{R}$ as
\eqn\bounh{H_{R} = {1\over{2m}} \int_{R} d^{2} {\bf x} \
{\left ( D_{i} \Psi \right )}^{\dag} {\left ( D_{i}
\Psi \right )}
-{1\over{2m}}\int_{R} d^{2} {\bf x}\ eB \rho\ ,}
where the subscript $R$ denotes integration of the radial
coordinate only up to $R$.
While being defined in the infinite plane, the new Hamiltonian mimics
the system in a disk of radius $R$.
In this finite geometry, the degeneracy of the lowest Landau level
is given by the magnetic flux in quantum units through the disk,
$\Phi\equiv R^2/\ell^2$.
In the plane and for integer filling $\nu=1$, $(\Phi +1)$ electrons
occupy approximatively a circular region of radius $R$ (see fig. 2).
We thus relate the number of particles $L+1$ to the radius $R$ as
follows
\eqn\defmu{\Phi = {R^{2} \over\ell^2} = L+\mu\ ,\qquad L\gg 1 \ ,\
\mu = O(1)\ .}

After this identification, let us show that $H_R$ correctly reproduces
the realistic spectrum descibed in the previous section.
Note that the eigenfunctions $\psi_{l}({\bf x})$
of the first Landau level in the plane
still diagonalize $H_{R}$ by angular integration.
The lowest Landau level Fock space is thus unchanged; however
the degeneracy is lifted and we obtain the one-particle spectrum
\eqn\espect{\varepsilon_{l}={\omega\over 2}\ {{\Phi^{l}
e^{-\Phi}}\over{l!}} \ (l-\Phi)\ .}
The field operator restricted to the first Landau level $\Psi$
in eq.\fllfo\ acquires a time dependence:
\eqn\flltd{\Psi({\bf x},t) = \sum_{l=0}^{\infty}\ a_{l}\
\psi_{l}({\bf x})
\ e^{-i\varepsilon_{l} t} \ .}

The spectrum in eq.\espect\ is shown in Fig. 6 for $\Phi =10$; it
becomes approximately linear
for $\Phi - \sqrt{\Phi} \le l \le \Phi + \sqrt{\Phi}$. The
minimum and maximum are located at $\Phi \pm \sqrt{\Phi}$,
respectively, and the maximal value of $\varepsilon_{l}$ is, in first
approximation, independent of $\Phi$ and given by
$\varepsilon_{\rm Max} \simeq 0.12 \omega$.
Since $\varepsilon_{\rm Max} \ll \omega$, the energies
of states in the first Landau level are well separated from the
energies of states in the second Landau level and, therefore,
the first Landau level retains its character\footnote{*}{
For filling $\nu=1$ we can neglect the mixing $O(\exp(-(R/\ell)^2))$
with higher Landau levels. }.
By using the Stirling formula, the spectrum \espect\
for\  \hbox{$L-\sqrt{L} \le l\le L+\sqrt{L}$} \ becomes
\eqn\espcm{\varepsilon_{l}\simeq {v\over R}
\ \left[  (l-L) - \mu \right]\ ,\qquad
v=\sqrt{eB\over{4\pi m^{2}}}\ ,}
and we recognize $v$ and $\mu$ as the velocity and chemical potential
parameters introduced in the previous section.
In this context, the ``light-velocity'' $v$ has to be considered
as a {\it phenomenological parameter}; its exact value depends
on the real confining potential felt by the electrons and
the value obtained in \espcm\ has not to be taken as a prediction.
The velocity will
appear as an overall parameter in the emerging conformal field
theory, whose universal results will be independent of its exact value.
Note also that our approach gives a vanishing $\beta_0$ and
is therefore appropriate to describe one edge of the annulus.
The gaps for charged excitations on the disk geometry will be
restored later in section 3.

The crucial point in our treatment of the confining pressure
is that the Hamiltonian $H_{R}$ in \bounh\ automatically reduces
to a $(1+1)$-dimensional Hamiltonian for the
{\it boundary} ($r=R$) {\it values} of the field operator \flltd.
To see that these constitute the {\it only degrees
of freedom} we employ an identity recently
pointed out in ref.
\ref\jpi{R. Jackiw and S.-Y. Pi, {\it Phys. Rev.} {\bf D42}
(1990) 3500.}:
\eqn\bogide{\left( D_{i}\Psi\right)^{\dag} \left(D_{i}\Psi\right)=
\left( D_{+}\Psi\right)^{\dag} \left(D_{+}\Psi\right)\ +
m\varepsilon^{ij}\partial_{i}J^{j}\ +\ eB\rho\ .}
Combining this with the self-duality condition \sdc\ for $\Psi$
restricted to the first Landau level, we see that $H_{R}$ reduces to
a pure boundary term given by
\eqn\boham{H_{R}\ =\ {1\over{4m}} \int_{0}^{2\pi R} dx\
{\Psi}^{\dag}(-i)\left(\partial_{x} - i{R\over\ell^2}\right) \Psi
+ c.c.\ ,}
where $x=R\theta$ is the 1-dimensional coordinate on the edge.
Note that the Hamiltonian density of this
``relativistic'' $(1+1)$-dimensional theory is the transverse
component of the non-relativistic current restricted to $r=R$.

We thus conclude that our modification of the original
Hamiltonian to $H_{R}$ given in \bounh\ corresponds to
introducing a confining potential due to the background;
the big advantage of our procedure is that it allows for an exact
analytic treatment. As a bonus, we obtain
an automatic reduction from a $(2+1)$-dimensional
non-relativistic Hamiltonian to a $(1+1)$-dimensional
``relativistic'' Hamiltonian. This reduction is the microscopic
analogue of the reduction in Chern-Simons
gauge theories
\ref\witten{E. Witten, {\it Comm. Math. Phys.} {\bf 121} (1989) 351.}.
\bigskip
\vfill\eject
\newsec{Thermodynamic limit and CFT on the cylinder}
\bigskip
\noindent{\bf 3.1 Thermodynamic limit}
\bigskip
We now study the thermodynamic limit
$L =(R/\ell)^2 -\mu\ \to \infty$ of the theory describing the
dynamics at one edge of the Hall sample.
In this limit, all quantities in the theory will acquire a manifest
$(1+1)$-dimensional relativistic form and we shall find conformal
invariance by explicit derivation of the Virasoro algebra.
In eq.\boham, we already derived the $(1+1)$-dimensional Hamiltonian.
However, it is not obvious that the field operator $\Psi(r=R)$ reduces
in the limit to the correct expansion for a relativistic theory.
Nonetheless this is indeed the case, as we now discuss.

We first notice that the theory defined by the Hamiltonian
\bounh\ contains both an ultraviolet cutoff $\ell$ and
an infrared cutoff $R$, which are tied together by the particle number
$L\sim (R/\ell)^2$.
The restriction to the linear range $|l-L|<\sqrt L$ of the
spectrum $\varepsilon_l$ in eq.\espcm~
amounts to the ultraviolet cut-off for the
one-dimensional momentum \hbox{$|p|=|(l-L)/R|<1/\ell \ $},
which becomes  irrelevant in the limit $R\to\infty$.
In this limit, the Fock space of the operators $a_l, a_l^{\dag} , $
for angular momentum states can be identified with the Fock space of a
relativistic $(1+1)$-dimensional fermion.
Similarly, the filled Landau level becomes the Dirac sea for the
relativistic fermion \stone.

To study the behaviour of the field operator,
it is convenient to shift variables as follows,
\eqn\shift{\eqalign{e_{l} &\equiv \varepsilon_{L+l}\ ,\cr
b_{l} &\equiv a_{L+l}\ , \cr
\Psi ({\bf x}) &= \left( {2\over \pi\ell^2} \right)^{1/4}\
e^{i(L+\mu)\theta} \ F_{R}^{\rm (cutoff)} ({\bf x})\ .\cr}}
When evaluated on the boundary, the field operator
$F_R^{\rm (cutoff)}$ acquires the form
\eqn\expf{\eqalign{ F_R^{\rm (cutoff)} \left(Re^{i\theta}, t \right)
&= \sum_{l=-L}^{\infty} {C_l\over\sqrt{2 \pi R}}\
e^{i(l-\mu) \theta }\ e^{-i e_l t}\ , \cr
C^2_l &=\left. {\sqrt{2\pi}\over{(L+l)!}}
\left({R\over\ell}\right)^{2L+2l+1}\ e^{-R^2/\ell^2}
\right\vert_{R^2 =(L+\mu)\ell^2} \ .\cr}}
In this expansion, the coefficients are approximated by
$C_l\sim \exp(-(l-\mu)^2/2L)$ for $L\to\infty$.
Since the width of this gaussian is $\sqrt{L}$, the $C_l$
give a smooth ultraviolet cutoff to the sum over $l$, which is limited
to the range $|l|<\sqrt{L}\simeq R/\ell$ of linear
energy \espcm. In this range, the wave functions have the correct
$(1+1)$-dimensional form which ensures orthonormality in the
Hilbert space on the circle.
Note that to leading order for $L\to\infty$, we can remove
this cutoff which would give subleading contributions
$O(1/\sqrt{L})=O(\ell/R)$.
(This will be explicitly checked in section $4$).
Therefore, we obtain the approximate field operator
\eqn\feff{ F_R \left(\theta , t \right)
= \sum_{l=-\infty}^{+\infty} {1\over{\sqrt{2 \pi R}}}\
e^{i(l-\mu) (\theta -{v\over R} t)}\ .}

In terms of the field $F_{R}$, the boundary Hamiltonian
introduced in the previous section, eq.\boham ,
takes the form
\eqn\boshi{H_{R}\ =\ {v\over 2} \int_{0}^{2\pi R} dx\
F^{\dag}_{R} \left( -i\partial_{x} \right) F_{R}
+ c.c.\ ,\quad x=R\theta  .}
Eqs.\feff\  and \boshi\  are the field and the
Hamiltonian of a relativistic chiral charged
fermion\footnote{*}{The chirality is determined by the sign
of the external magnetic field $B$.}
(Weyl fermion), moving with ``light-velocity'' $v$ on a circle of
circumference $2\pi R$.

A non-trivial aspect of the thermodynamic limit is the
role of the chemical potential $\mu$.
Eq.\feff~ allows a continuous spectrum of
boundary conditions
\eqn\boundary{F_R(\theta +2\pi) =\  e^{-i2\pi\mu} \ F_R(\theta)\ }
parametrized by the fractional part of $\mu$.
Such general boundary conditions are possible for the Weyl fermion
\gins, and affect the spectrum of the theory, as we shall see
later. The two cases mostly considered in  string theory are
called:

{\it i)} $\mu=0$: Ramond (R),  or periodic b.c. on the circle;

{\it ii)} $\mu={1\over 2}$: Neveu-Schwarz (NS), or anti-periodic
b.c. on the circle.

\bigskip
\bigskip
\noindent
{\bf 3.2 Conformal Field Theory on the Cylinder}
\bigskip
Conformal transformations in $(1+1)$ dimensions are analytic
reparametrizations of one complex coordinate.
A distinctive feature of a conformal invariant theory
is the possibility to express all quantities as analytic
functions of this complex coordinate, call it $\eta$ \bpz.
This variable parametrizes an Euclidean space-time plane.
Our theory, given by eqs.\feff~ and \boshi\  is defined
on the Minkowskian cylinder $(R\theta,t)$ made
by the circle and the time coordinate.
By rotating to Euclidean time $t=-i\tau$, we identify
the conformal coordinate as
\eqn\mapping{\eta = e^{{1\over R} (v\tau +i R\theta)} \ ,\qquad
{\rm Arg}\  \eta = {\rm Arg}\  z \ ,}
where $z$ parametrizes the original plane of the electrons.
This mapping relates the cylinder with the conformal
plane $\eta$, such that its unit circle $|\eta|=1$ corresponds
to the edge $|z|=R$ (see Fig. 7).

The field operator \feff~ is anti-holomorphic in this variable,
\eqn\feta{F_R({\bar\eta})\ =\
{{\bar\eta}^{1\over 2} \over \sqrt{2\pi R}}\
\sum_{l=-\infty}^{+\infty}
{\bar\eta}^{-l +\mu -{1\over 2}}\ b_{l}\
={{\bar\eta}^{1\over 2} \over \sqrt{2\pi R}}\
F(\bar\eta) \ .}
Similarly, one finds
\eqn\fdageta{F_R^{\dag}({\bar\eta})\ =\
{{\bar\eta}^{1\over 2} \over \sqrt{2\pi R}}\
\sum_{l=-\infty}^{+\infty}
{\bar\eta}^{l -\mu -{1\over 2}}\ b^{\dag}_{l}\
={{\bar\eta}^{1\over 2} \over \sqrt{2\pi R}}\
F^{\dag} ({\bar\eta}) \ .}
In these equations, we stripped off an additional term in order to make
contact with the standard notation of CFT \gins.
The equations of motion now coincide with the analyticity condition
$\partial_\eta F=\partial_\eta F^{\dag} = 0$, and the
Hamiltonian and $U(1)$ current densities are also chiral
\eqn\dens{\eqalign{
{\rho}_{R}(\bar\eta) &=\ F^{\dag}_{R} F_{R} \ ,\cr
H_R ={1\over i} \oint d{\bar\eta}\ {\cal H}_R(\bar\eta), \quad
{\cal H}_{R}(\bar\eta) &
={v \over 2} \left( \partial_{\bar\eta} F^{\dag}_{R} F_{R}
\ -\ F^{\dag}_{R} \partial_{\bar\eta} F_{R}\ \right) \ .\cr}}
The following table clarifies the meaning of these two densities
in $(2+1)$ and $(1+1)$ dimensions.
$$\eqalign{
\quad \quad \underline{{\rm Non-rel.}\ (2+1)}&\qquad\qquad
                                    \underline{{\rm Rel.} \ (1+1)}\cr
\rho_R \qquad {\rm charge\ \  density} &\qquad\qquad
                                    {\rm chiral \ \ current}  \cr
{\cal H}_R \quad{\rm transverse\ \ current} &\qquad\qquad
                         {\rm stress\ \  tensor}\ {\bar T}(\bar\eta)\cr}
$$

We now compute the chiral algebra of this
field theory. To this end, we define the charges
\eqn\mode{\eqalign{ {\rho}_{n} &\equiv
{R\over i}\ \oint \ d\bar\eta\ {\rho}_{R}(\bar\eta)\ {\bar\eta}^{n-1}, \cr
L_{n} &\equiv {R\over iv}\
\oint \ d\bar\eta\ {\cal H}_{R}(\bar\eta)\ {\bar\eta}^{n}. \cr}}
The charges $\rho_n$ and $L_n$ generate
local gauge and conformal transformations. Global transformations
are generated by $\rho_0$ and $\{ L_{-1} ,L_0 , L_1 \}$.
Note that the total charge operator and the
Hamiltonian of the $(1+1)$-dimensional relativistic theory
are given by
$\rho_0$ and $H_R ={v\over R} L_0$, respectively.
The dimensionless Hamiltonian $L_0$ of CFT will describe
the $1/R$-dependence of the spectrum of edge excitations.
By using the field expansions \feta\fdageta\ in eq.\mode, we obtain
\eqn\modes{\eqalign{ {\rho}_{n} &=\ \sum_{l=-\infty}^{+\infty}
\ b^{\dag}_{l-n}b_{l}\ , \cr
L_{n} &=\ \sum_{l=-\infty}^{+\infty}\ (l-{n\over 2} -\mu)
\ b^{\dag}_{l-n}b_{l}\ . \cr}}
Clearly, $\rho_0$ and $L_0$
have to be
regularized with some normal ordering prescription. To this end,
we note that, with the notation introduced in \shift , the
original filled Landau level state $|\Omega\rangle$ is defined by
\eqn\vcond{\eqalign{ b_l |\Omega\rangle &=\ 0,\ l>0\ , \cr
b^{\dag}_l |\Omega\rangle &=\ 0,\ l\le 0\ , \cr}}
and thus becomes the filled Dirac sea for the relativistic
field theory of edge excitations.
The normal ordering prescription for this ground state
is obtained by writing the annihilation
operators $b_l$ ($l>0$) and $b^{\dag}_{l}$ ($l \le 0$)
to the right-hand side of the creation operators
$b^{\dag}_{l}$ ($l> 0$) and $b_l$ ($l\le 0$).
This prescription amounts to an infinite subtraction,
therefore finite normal ordering constants should be included in
$\rho_0$ and $L_0$, and they are fixed by algebraic conventions as
follows.

The normal-ordered charges $\rho_n$ and $L_n$ satisfy the following
chiral algebra
\eqn\chalg{\eqalign{[{\rho}_{n},{\rho}_{m}] &=\ n\ \delta_{n+m,0}
\ , \cr
[L_{n},{\rho}_{m}] &=\ -m\ \rho_{n+m}\ , \cr
[L_{n},L_{m}] &=\ (n-m)\ L_{n+m}\ +\ {c\over 12}\ (n^{3} - n)\
\delta_{n+m,0}\ ,\quad c=1\ . \cr}}
(The explicit computation is described for example in ref.
\ref\itzy{C. Itzykson and J. M. Drouffe, {\it ``Statistical Field Theory''},
Cambridge Univ. Press, Cambridge (1989), vol. 2, pages 540 ff.}).
The first algebra in \chalg\ is the {\it U(1) Kac-Moody algebra} for
the generators $\rho_n$ of local gauge transformations;
the third algebra is the {\it Virasoro algebra} for the generators
$L_n$ of local conformal transformations.
As is well known, the central extensions in
these two algebras come from quantum effects associated with the
infinite Dirac sea in the thermodynamic limit. The coefficient of $n^3$
in the Virasoro algebra does not depend on the finite normal-ordering
constants and determines the central charge $c$. The latter is found
to be $c=1$, as it should be for a {\it Weyl fermion}.
Instead, the coefficient of the linear term in the
Virasoro central extension can be shifted by redefinitions of
$L_0$ by a constant. The convention in \chalg\ is standard and
fixes the value of $L_0 |\Omega\rangle$. The value of
$\rho_0 |\Omega\rangle$ is fixed by the requirement of the
absence of an anomaly in the mixed commutator. Altogether,
the properties of the ground state are summarized by
\eqn\vacpr{\eqalign{ L_n |\Omega,\mu\rangle\ &=\ 0\ ,\qquad \qquad
\qquad \qquad \qquad
\rho_n |\Omega,\mu\rangle\ = 0\ ,\ n>0
\ , \cr
L_{0} |\Omega,\mu\rangle\ &= \left({1\over 8}+{{\mu^{2}-\mu}\over 2}
\right) |\Omega,\mu\rangle\ ,\qquad \rho_0 |\Omega,\mu\rangle\ =
\left( {1\over 2} - \mu \right) |\Omega,\mu\rangle\ . \cr}}
Note the dependence of the ground state on the chemical potential $\mu$.
In the thermodynamic limit, the filled Landau level is a highest-weight
state of the ($c=1$) chiral algebra \chalg\ with charge $Q_0={1\over 2}
-\mu$ and conformal dimension $h_0={1\over 8}+{1\over 2}({{\mu}^{2}-
\mu})$.
In the next section, we shall determine the value of $\mu$.
Given that $|\Omega,\mu\rangle$ represents the original filled Landau
level state, we recognize \vacpr\ as the {\it algebraic
characterization of the incompressibility} of this state.
It can be shown
\ref\ctznext{A. Cappelli, C. A. Trugenberger and G. R. Zemba,
{\it to appear}.}
that the highest-weight conditions \vacpr\  are equivalent to
the $W_\infty$-symmetry conditions recently discussed in ref.\ctz.

Local conformal and gauge transformations, generated by the charges
$L_n$ and $\rho_m$, are {\it dynamical symmetries} of the theory.
This is reflected in the fact that the charges $L_n$ and $\rho_m$
do not commute with the Hamiltonian $L_0$; this non-commutativity
is, however, compensated by the explicit time-dependence of the
charges $L_n$ and $\rho_m$. Indeed, this is such that
\eqn\dynsy{\eqalign{ {{dL_n}\over{d\tau}} &=\ {{\partial L_n}\over
{\partial\tau}}\ +\ {v\over R}[L_0 , L_n ]\ =0\ , \cr
{{d\rho_n}\over{d\tau}} &=\ {{\partial \rho_n}\over
{\partial\tau}}\ +\ {v\over R}[L_0 , \rho_n ]\ =0\ . \cr}}
This is sufficient to guarantee that all states in the theory fall
into highest-weight representations of the chiral algebra \chalg ,
characterized by the eigenvalues $h$ of $L_0$ (conformal dimensions)
and $Q$ of $\rho_0$ (charge). Such states are generated from the
ground state by the action of {\it primary fields} $V_{h,Q}$: these
specify the operator content of the CFT \bpz\gins.
\bigskip
\bigskip
\noindent
{\bf 3.3 Operator content of the CFT}
\bigskip
To extract the operator content of our theory, we compute the
grand-canonical partition function on the torus obtained by
compactifying the Euclidean time,
\eqn\grandc{\eqalign{
{\cal Z}(q,w)\ & \equiv \ {\Tr}\left( e^{-TH_R} \ w^{\rho_0}\right)
=\ {\Tr}\left( q^{L_0 -{1\over 24}} w^{\rho_0}\right)\  , \cr
q\ &\equiv\exp\left( -{vT\over R} \right) ,\cr}}
where $T$ is the Euclidean time period and $w$ the fugacity.
The factor $1\over 24$ in the exponent is the Casimir energy which
will be discussed in detail in the next section.
${\cal Z}(q,w)$ is readily
computed with standard statistical mechanics techniques, once we
realize that our relativistic theory describes ``electrons''
with spectrum $(n-\mu)$ $(n \ge 1)$, and ``positrons'' with
spectrum $(n -(1-\mu))$ $(n \ge 1)$, as is clear from \modes :
\eqn\gcpf{ {\cal Z}(q,w)\ =\ q^{{1\over 12} +
{1\over 2}({\mu}^{2} - \mu)}
\ w^{{1\over 2}-\mu}\ \prod_{n=1}^{\infty} \left( 1+ w\ q^{n-\mu}
\right)\ \left(1+w^{-1}\ q^{n-(1-\mu)}\right)\ .}
This expression can be further simplified by using the following
version of the Jacobi triple product identity \tables :
\eqn\jtpi{ q^{{1\over 2}\left( {1\over 2} - \mu \right)^{2}}
\prod_{n=1}^{\infty}\left(1-q^{n}\right)\
\left( 1+ w\ q^{n-\mu} \right)
\ \left(1+w^{-1}\ q^{n-(1-\mu)}\right)\ =\ \sum_{n=-\infty}^{\infty}
q^{{1\over 2}\left( n+{1\over 2} - \mu \right)^{2}}\ w^{n}\ .}
Indeed, use of this identity leads to the final result:
\eqn\partfn{ {\cal Z}(q,w)\ =\ {1\over{\eta(q)}}
\ \sum_{n=-\infty}^{\infty}
q^{{1\over 2}\left( n+{1\over 2} - \mu \right)^{2}}\ w^{n+{1\over 2}-
\mu}\ ,}
where $\eta(q)$ denotes the Dedekind funtion
\eqn\dedek{ \eta (q)\ =\ q^{1\over 24}\ \prod_{n=1}^{\infty}
\left(1-q^{n}\right)\ .}
The partition function contains two basic factors. The factor
$1/\eta (q)$ represents the contribution from neutral particle-hole
excitations across the Fermi level; the second factor is
a sum over charged sectors which are in one-to-one correspondence
with the primary fields $V_{h_n,Q_n}$\ .
The exponents of $q$ and $w$ give their conformal dimensions
\footnote{*}{Degenerate representations for $c=1$, having dimensions
$h=k^{2}/4$, $k \in {\bf Z}$, never arise here.}
and charges, respectively:
\eqn\handq{\eqalign{ h_{n}\ &=\ {1\over 2}\left(n+{1\over 2} -\mu
\right)^{2}\ , \cr
Q_{n}\ &=\ n+{1\over 2} -{\mu}\ . \cr}}

Note that from the point of view of the CFT the integer part of $\mu$
is irrelevant since it can be reabsorbed by a shift of the summation
index $n$ in eq.\partfn.
Only the fractional part of $\mu$ is relevant, since it determines
the boundary conditions.
Actually, we are now going to argue that $\mu$ is dynamically fixed
to be $1/2$.

To this end we first note that the confining potential and the
non-degenerate spectrum $\varepsilon_l$ arise from the
neutralizing background.
Moreover, as we show in the next section, the renormalized charge
operator $\rho_0$ is the physical charge, since it correctly couples to
additional electro-magnetic fields, thereby leading to the measured
Hall current.
Therefore, the ground state should be a neutral eigenstate of
$\rho_0$. We thus conclude that $\mu$ dynamically self-tunes
to the value
\eqn\muns{ \mu\ =\ {1\over 2}\ .}
This implies that the ground state
$|\Omega\rangle \equiv |\Omega,{\mu ={1\over 2}}\rangle$
is $SL(2,{\bf C})$ invariant, since
it is annihilated by all three generators of
global conformal trasformations,
$ L_{k}|\Omega\rangle =0\ , \ k=-1,0,1\ $.
It implies also the absence of spin fields
\ref\gsw{M. B. Green, J. H. Schwarz and E. Witten, {\it Superstring
Theory}, Cambridge Univ. Press, Cambridge, (1988).}
from the spectrum of the CFT.

With $\mu=1/2$,
the conformal dimensions and charges of the primary fields reduce to
\eqn\haqsl{\eqalign{ h_{n}\ &=\ {1\over 2} n^{2}\ , \cr
Q_{n}\ &=\ n\ . \cr}}
Every one of these primary fields can be obtained by operator product
expansion of the two basic fields $V_{{1\over 2},-1}\equiv F$ and
$V_{{1\over 2},+1}\equiv F^{\dag} $, where $F,F^{\dag} $
are the fermion fields previously introduced in eqs. \feta,\fdageta.
We thus conclude that the theory describing the dynamics of the
edge excitations in the thermodynamic limit is the $c=1$ CFT
of a Weyl fermion with Neveu-Schwarz boundary conditions.

The Hall system on the annulus geometry can be described by
taking a copy of the CFT
for each boundary. The one-particle spectrum for the annulus of radii
$R_1, R_2$ described in section 2 (see Fig. 4) can be reproduced in
our approach by the Hamiltonian
\eqn\hannulus{H_{R_1,R_2} \ = \ {1\over 2m}
\int_{R_1 < r < R_2} \ d^2 {\bf x}
\left( \left(D_i \Psi\right)^{\dag} \left(D_i\Psi\right) - B\rho \right)
\ =\ H_{R_2} -H_{R_1} \ ,}
where $H_{R_1}$ and $H_{R_2}$ are two copies of the previously
discussed Hamiltonian.
For \hbox{$R_2-R_1 \to\infty$}, the one-dimensional Hilbert spaces at
each edge are coupled only by overall charge conservation.
The total partition function is thus given by
\eqn\zannulus{ \eqalign{ {\cal Z}_{\rm annulus}\ =&\
{1\over \eta (q)\eta(\bar q)}\ \sum_{n=-\infty}^{+\infty}
\left( q \bar q \right)^{{1\over 2}n^2} \ w^n \ ,\cr
q\equiv & \exp\left( -{v_2 T\over R_2} \right)\ , \qquad
{\bar q} \equiv  \exp\left( -{v_1 T\over R_1} \right) \ .\cr} }
While neutral excitations at the two edges remain independent, charged
excitations require the transfer of particles from one edge to the
other.
\bigskip
\bigskip
\noindent{\bf 3.4 Deformed CFT on the disk geometry}
\bigskip
As emphasized in section 2, on a disk geometry the charged excitations
acquire a gap that spoils conformal symmetry.
Nonetheless, much information can be extracted from the CFT
derived in the preceding sections. The crucial point is that
$\rho_0$ is a Casimir of the chiral algebra \chalg.
Therefore one can modify the Hamiltonian by adding to it any
function of $\rho_0$ without changing the Hilbert space of the theory.
We can add the gaps by hand with the following redefinition:
\eqn\hhat{ H_R \ \to {\hat H}_R \ = \ {v\over R} L_0
+\gamma (\rho_0) \ ,\qquad\quad
\gamma (n) = \cases { n\beta_0, & $n \ge 0\ $, \cr
                      n(\Delta - \beta_0 ), & $ n<0$ \ .\cr} }
In any irreducible representation of the chiral algebra with
charge $n$ the additional piece is a multiple of the identity.
The infinite charges $\rho_n$ and $L_n$ are still (dynamically)
conserved with respect to the modified Hamiltonian $\hat H_R$
as in eq.\dynsy.
Therefore it still makes sense to label states of the Hilbert space
according to the eigenvalues of these charges.
The resulting {\it deformed} CFT is fully consistent.
Its partition function is given by
\eqn\zdisk{ {\cal Z}_{\rm disk} = {1\over \eta (q)} \
\sum_{n=-\infty}^{+\infty}
q^{ {1\over 2}n^2 + {R\over v}\gamma (n)} \ \ w^n \ .}
The ``conformal part'' $vL_0/R$ of the Hamiltonian $\hat H_R$
determines the universal $1/R$-corrections to the non-universal gaps
$\beta_0$ and $(\Delta -\beta_0)$.

Note that the presence of gaps on the disk geometry has been somewhat
neglected in the literature on edge excitations.
In particular, the effective Chern-Simons theory approach applies, strictly
speaking, only to fully gapless geometries like the annulus.
\bigskip
\vfill\eject
\newsec{ Integer Quantum Hall Effect: the CFT point of view}
\bigskip
In this section, we shall explore the implications of the
CFT derived in the previous section for the physics of the
$\nu =1$ quantum Hall effect. Clearly, this effective theory
describes only some properties of the quantum Hall states;
however, conformal invariance implies that these properties are
{\it universal}.
\bigskip
\noindent{\bf 4.1 Finite-size spectra on the disk geometry}
\bigskip
In section 3 we have explained how all quantities of the
CFT on the cylinder can be obtained from the corresponding quantities
on the plane $\eta$ by the conformal map \mapping. In the case of
the Hamiltonian, however, we did not pay
attention to normal ordering effects in eq.\mode.
The correct Hamiltonian \cardy\  on the cylinder is given by
\eqn\casen{H_R\ =\ {v\over R}\left(L_0 -{c\over 24}\right)\ ,
\qquad (c=1).}
The additional term depends on the central charge of the theory
and is called the {\it Casimir energy}.
It arises from the mismatch of normal ordering procedures on the plane
and on the cylinder.

Suppose we have computed the ground-state energy of the $\nu=1$
quantum Hall state in the disk geometry of radius $R$.
The finite-size expansion of this energy will be generically
\eqn\ezero{E_0\ =\
{R^2\over \ell^3} \alpha_2 +\  {R\over \ell^2} \alpha_1 +\
{1\over \ell} \alpha_0 +\      {1\over R} \alpha_{-1} +\
{\ell\over R^2} \alpha_{-2} +\dots \ .}
In this equation, $\ell$ is the magnetic length
(the ultraviolet cutoff) and
$\alpha_n$ are dimensionless coefficients that can depend on
$\ell$, the electron mass $m$ and other
parameters of the microscopic Hamiltonian.
The leading term gives a finite energy per particle since $(R/\ell)^2$
is the number of electrons. The other terms are subleading in the
experimental regime of large flux through the sample,
$\Phi= (R/\ell)^2 \gg 1$.
On the other hand, the result of the CFT is given by
eq.\casen\ combined with $L_0|\Omega\rangle= 0$:
\eqn\eground{E_0 \ = \ \langle\Omega | H_R |\Omega\rangle \ =
\ - {v\over 24} {1\over R} \ .}
Actually, in the CFT the terms
$\alpha_2$ , $\alpha_1$ and $\alpha_0$ have been put to zero
by the infinite subtraction of normal ordering.
The terms $\alpha_{-2},\alpha_{-3},\dots$ represent the
higher order corrections for $R\to\infty$ neglected in the
thermodynamic limit of section $3$.
Therefore, CFT leads to the prediction $\alpha_{-1} = -{v\over 24}$.
The discussion of section $3$ shows that this prediction is
universal because it is independent of the details
of the confining potential.
The velocity $v$ is a free parameter; it can be determined from the
spectrum of excitations as we discuss hereafter.
Note that the CFT prediction is a subleading correction
to the bulk properties discussed by Laughlin \laugha\
and verified in boundaryless geometries
\ref\sphere{F. D. M. Haldane and E. H . Rezayi, {\it Phys. Rev. Lett.}
{\bf 54} (1985) 237;
G. Fano, F. Ortolani and E. Colombo, {\it Phys. Rev. } {\bf B34}
(1986) 2370.}.

The Casimir energy \casen\ implies a universal term
\eqn\press{P\ =\ -{v\over{48\pi}}{1\over R^{3}}\ ,}
in the pressure $P$ of the electron
liquid. An independent measurement of the electron liquid pressure
and the effective velocity $v$ of edge excitations, might thus
provide an experimental verification of the CFT.
Another possibility is the measurement of the contribution of
the Casimir energy to the specific heat \wen.

In order to illustrate the finite-size expansion
\ezero, let us compute the
ground state energy $E_0$ directly by using the spectrum \espect .
This gives
\eqn\enne{E_0\ =\ \sum_{l=0}^{L}\ {\omega \over 2}\ {{{\Phi}^{l}
e^{-\Phi}}\over{l!}}\ (l-\Phi)\ =\ -{\omega \over 2} e^{-\Phi}\
{{\Phi}^{L+1} \over{L!}}\ .}
Using $\Phi =(R/\ell )^2= L + {1\over 2}\quad (\mu={1\over 2})$,
and Stirling's formula
\eqn\stirl{L!\ =\ L^{L}\ e^{-L}\ {\sqrt{2\pi L}}\ \left(
1+{1\over{12L}} +O\left({1\over L^2}\right) \right)\ ,}
we obtain the expansion of $E_0\ $:
\eqn\enstir{E_0\ =\ - v\ {R\over{{\ell}^2}}\ -\ {v \over 24}{1\over R}
\ +O\left( {\ell\over R^2}\right) \ .}
As expected, the $1/R$-term is in perfect agreement with the
prediction of the CFT.

A further result of the CFT is that {\it excitations}
about the incompressible filled Landau level state can all be
classified according to representations of the chiral
algebra \chalg . As we discussed in section 3, there are two
types of such excitations: neutral particle-hole excitations
(edge excitations), and charged excitations.
The charged excitations are in one to one
correspondence with the primary fields of the chiral algebra, and
are identified with the {\it quasi-hole} and {\it quasi-particle}
excitations of Laughlin's theory. The spectrum of excitations
can be extracted from the partition function \zdisk\ :
\eqn\qhqp{\Delta E_{n,k}\ \equiv\ E_{n,k} - E_0 \ = \
\gamma (k) \ +\ {vh_{k} \over R}\ +\
{vn\over R}\ ,}
with $n,k \in {\bf Z}$ and $0<n\ll \sqrt{L}$ .
The second term is the universal
$1/R$-correction to the gap and is determined by the
conformal dimension $h_k =k^2 /2$ of the primary fields.
Conformal invariance implies that this is the same for
quasi-particles and quasi-holes.
The third term is the spectrum of neutral edge excitations
in any given charged sector labelled by $k$ (the conformal family).

Numerical simulations in the disk geometry were carried out in refs.
\ref\morf{R. Morf and B. I. Halperin, {\it Phys. Rev. } {\bf B33}
(1986) 2221;
N. Datta and R. Ferrari, {\it Investigations of the
Laughlin State}, Max-Planck preprint MPI-Ph/92-16.}
\ref\stonedge{M. Stone, H. W. Wyld and R. L. Schult, {\it ``Edge Waves
in the Quantum Hall Effect and Quantum Dots''}, preprint
ILL(TH)-91-27.}
\wenrev .
The authors of refs.\stonedge \wenrev\ concentrated on neutral
edge excitations in systems of interacting electrons.
These studies show a good agreement with the multiplicities
predicted by conformal invariance (the $1/\eta$-term in the partition
function \zdisk ), but did not yet verify the linear relativistic
spectrum.
Finite-size effects in the spectrum were addressed in ref.\morf ;
unfortunately, the numerical precision is not yet sufficient to allow
a quantitative comparison with our predictions.
It would be most interesting to improve these analyses.
\bigskip
\bigskip
\noindent{\bf 4.2 Monodromies and statistics of excitations}
\bigskip
An important consequence of CFT is that it predicts
the monodromies of the primary fields. Through analytic continuation,
these determine then the {\it statistics} of the quasi-particle and
quasi-hole excitations.
The identification between CFT correlators and physical wave functions
is done at $t=0$, and involves therefore only static properties of
the CFT. The deformation of the CFT discussed in the previous section
is not relevant for the following discussion of monodromies.

Let us compute the conformal correlators of two primary fields
at points ${\bar \eta}_1$ and ${\bar \eta}_2$:
\eqn\correl{\eqalign{\langle\Omega| b^{\dag}_0 b^{\dag}_{-1}
F({\bar \eta}_1) F({\bar \eta}_2)|\Omega\rangle\ &=\
({\bar \eta}_1 - {\bar \eta}_2)\ , \cr
\langle\Omega| b_1 b_2 F^{\dag}({\bar \eta}_1) F^{\dag}({\bar \eta}_2)
|\Omega\rangle\ &=\ ({\bar \eta}_1 - {\bar \eta}_2)\ . \cr}}
Note that charge conservation requires the ground state of charge
$\mp 2$ on the left-hand side of these expressions.
By applying the conformal map \mapping , and restoring the factors
extracted in \shift\feta\fdageta , we obtain the wave-functions of the
corresponding quasi-hole and quasi-particle excitations, located
in the physical plane at
$z_1 = R e^{i\theta_1}$ and $z_2 = R e^{i\theta_2}$ :
\eqn\corrqp{\eqalign{\langle\Omega| b^{\dag}_L b^{\dag}_{L-1}
\Psi(R e^{i\theta_1}) \Psi(R e^{i\theta_2})|\Omega\rangle\ &=\
{1\over{{\sqrt{2{\pi}^3}} R \ell}}\ e^{i(L-1)(\theta_1 + \theta_2 )}
\ (e^{i\theta_2} - e^{i\theta_1})\ , \cr
\langle\Omega| b_{L+1} b_{L+2} {\Psi}^{\dag}(R e^{i\theta_1})
{\Psi}^{\dag}(R e^{i\theta_2})|\Omega\rangle\ &=\
{1\over{{\sqrt{2{\pi}^3}} R \ell}}\ e^{-i(L+1)(\theta_1 + \theta_2 )}
\ (e^{-i\theta_1} - e^{-i\theta_2})\ . \cr}}
It is easy to check that this is exactly the result one would get
by direct computation of the same quantities using the original field
operator of the Landau levels. In particular, the
monodromies \correl\ determine the analytic parts
$(z_1 z_2)^{L-1} (z_2 - z_1 )$ and
$({\bar z}_1 {\bar z}_2 )^{L+1} ({\bar z}_1 - {\bar z}_2) \ $
of the quasi-hole and quasi-particle wave-functions and,
therefore, also the {\it fermionic statistics} of these excitations.
The fact that we are dealing with a $c=1$ CFT guarantees
the {\it polynomial character} of multi-excitations wave-functions;
this ensures in turn that these excitations carry {\it Abelian
statistics}.

The above computation clarifies the formal analogy between conformal
correlators and Laughlin's wave-functions pointed out by several
authors \fub\dlt\napo\moore.
The conformal correlators are defined only for $r=R$.
However, by analytic continuation, they determine the analytic part
of the excitation wave-functions of the full $(2+1)$-dimensional
non-relativistic theory. Therefore, the correlators of the
CFT are indeed sufficient to fix the statistics of the excitations.
However, we stress that the main relation is between conformal
correlators and excitation wave-functions, not Laughlin's
ground-state wave-functions.
\bigskip
\bigskip
\noindent{\bf 4.3 Long-range order on the boundary}
\bigskip
Conformal invariance dictates the form of the two-point function
of primary fields of conformal dimension $h$.
This has the power-law behaviour
\eqn\twopo{\langle\Omega|F^{\dag}({\bar \eta}_1) F({\bar \eta}_2)
|\Omega\rangle\ =\ {1\over{({\bar \eta}_1 -
{\bar \eta}_2)^{2h} }}\ ,\qquad h={1\over 2} \ \ .}
We can easily obtain the corresponding two-point function of the
CFT on the cylinder by applying the conformal map \mapping\
and restoring factors as in \corrqp.
We thus obtain the {\it density matrix} for points on the boundary:
\eqn\twopom{
\langle\Omega |{\Psi}^{\dag}(Re^{i\theta_1})
\Psi (Re^{i\theta_2})\ |\Omega\rangle\
= \ \sqrt{2\over\pi \ell^2} \ {1\over 2\pi R}\
e^{iL(\theta_2 -\theta_1)}\
{e^{i\theta_2} \over e^{i\theta_2} - e^{i\theta_1 } } \ .}
This implies that the filled Landau level state $|\Omega\rangle$
supports {\it algebraic long-range order on the boundary}, i.e.
the absolute value of the density-matrix for points on the boundary
at $r=R$ is proportional to the inverse of their distance (in the
$(2+1)$-dimensional sense). This long-range order is characterized
by the exponent $2h=1$, $h$ being the conformal dimension of the
primary fields entering the density-matrix. In Appendix A, we check
that this non-trivial result can be also obtained directly
from the Landau level expression of the density matrix
by a rather involved computation.
This confirms further the validity of the predictions of the CFT.
\bigskip
\bigskip
\noindent{\bf 4.4 Hall current in the annulus geometry}
\bigskip
We now derive the Hall current from the $(1+1)$-dimensional point
of view. To this end we consider the theory at one edge of the
annulus, say the outer one, and we add a flux tube at the center:
\eqn\delb{\delta B\ =\ 2\pi\ \Phi_0\ \delta^{(2)}({\bf x})\ .}
In our circular geometry, the appropriate
choice of gauge to describe this additional magnetic field is
\eqn\delpot{\delta A_i \ =\ \Phi_0\ \epsilon^{ij}\ {x^{j}\over{
|{\bf x}|^{2} }}\ .}
Accordingly, the one-particle Hamiltonian \elec\ is modified to
\eqn\delh{H\ =\ -{1\over{2m}}\ \left( {\bf \nabla}
- i e({\bf A}+ \delta {\bf A}) \right )^{2} \
- {e \over{2m}} (B+\delta B)\ ,}
The normalized energy and angular momentum eigenstates of this new
problem are
\eqn\eig{{\hat \psi}_{n,l}\ =\ e^{ie\Phi_0 \theta}\ \psi_{n,l}\ ,}
where $\psi_{n,l}$ are the Landau level wave-functions \psdhll .
While the energy eigenvalues $\omega_n$ corresponding to
${\hat \psi}_{n,l}$ are unchanged, the angular momentum eigenvalues
are modified to $(l+e\Phi_0)$. Following all the steps that led to
\feff\ and \boshi ,we find the following modifications due to the
additional flux tube $\delta B$:
\eqn\modif{\eqalign{H_{R}\ &=\ {v\over 2} \int_{0}^{2\pi R} dx\
{\hat F}^{\dag}_{R} \left( -i\partial_{x}+\ e\delta A_x \right)
{\hat F}_{R}+ c.c.\ ,\cr
\delta A_{x}\ &=\ -{{\Phi_0}\over R}\ \qquad , \qquad
{\hat F}_{R}\ =\ e^{ie{\Phi_0 \over{R}} x}\ F_R\ .\cr}}

We now allow $\Phi_0$ to vary in time; this corresponds to a transverse
electric field
\hbox{$E^{i} \equiv \partial_t A_i = {\dot \Phi_0} \epsilon^{
ij} x^{j} |{\bf x}|^{2}$,} which reduces to a $(1+1)$-dimensional
uniform {\it electric field}
\eqn\elec{E\ =\ \partial_t \delta A_{x}\ =\ -{{\dot \Phi_0}\over R}\ ,}
on the boundary at $r=R$. The effective theory \modif\ describes
therefore a $(1+1)$-dimensional Weyl fermion coupled to an
external electric field. As it is well-known, such a theory is
affected by the {\it chiral anomaly}
\ref\tjzw{For a review see: S. B. Treiman, R. Jackiw, B. Zumino
and E. Witten, {\it Current Algebra and Anomalies,}
Princeton Univ. Press, New Jersey (1985).},
which amounts to a violation of charge conservation
due to the external electric field. In the formulation \modif, the origin
of this phenomenon is hidden in the time-dependence of the normal
ordering procedure. However, the anomaly can be exposed by performing
a time-dependent transformation to the $\delta A_x = 0$ gauge. In
this gauge, we recover the previous field operator $F_R$; the price
for this is, however, an additional piece proportional to the density
in the Hamiltonian:
\eqn\modaf{\eqalign{H_{R}\ &=\ {v\over 2} \int_{0}^{2\pi R} dx\
F^{\dag}_{R} \left(-i \partial_{x} \right)
F_{R}+ c.c.\ +\ \int_{0}^{2\pi R}\ dx\ e\delta A_0\ \rho_R \ ,\cr
\delta A_{0}\ &=\ {{\dot \Phi_0}\over R} x\ \qquad , \qquad
E\ =\ -\partial_x A_0\ . \cr }}
We can now compute the time-variation of the normal-ordered density
$\rho_R$ by simply commuting it with the Hamiltonian $H_R$. The
commutator of the first term in $H_R$ with $\rho_R$ is fixed by the
second algebra in \chalg\ and gives $(-v\partial_x \rho_R)$; the
commutator of the additional term in $H_R$ with $\rho_R$ is determined
by the first algebra in \chalg . It is non-vanishing due to the central
extension in this Kac-Moody algebra and leads to the anomaly
equation
\eqn\anom{(\partial_t\ +\ v\partial_x) \rho_R\ =\ {e\over{2\pi}}\ E\ .}
This equation tells us the rate of production of particles
by the external electric field: $\partial_t Q = e E$. From
the $(1+1)$-dimensional point of view, this charge pops out from the
Dirac sea (spectral flow picture), which is an infinite reservoir.
The identification of the Dirac sea with the quantum
Hall droplet allows then to interpret the anomaly as a radial flow
of charge through the boundary at $R$, {\it i.e.}, a {\it Hall
current}
\eqn\hallcu{J^{i}\ =\ {e\over{2\pi}}\ \epsilon^{ij} E^{j}\ .}
Note however that in the $(2+1)$-dimensional non-relativistic theory
there is no net anomaly.
The charge flowing through the outer boundary is provided
by the inner boundary, and thus the anomalies of the CFTs at
the two edges cancel each other.
\bigskip
\vfill\eject
\newsec{Chiral boson and its canonical quantization}
\bigskip
It is well known that a Weyl fermion is equivalent to a free chiral
boson \gsw. The free fermion and the free boson give two equivalent
representations of the chiral algebra discussed above.
However, the free boson theory is more convenient because it can also
represent other $c=1$ CFTs, which correspond to interacting fermions.
In the next section, we shall apply these to the fractional Hall
effect.

There are two basic approaches to studying chiral $c=1$ CFTs.
One is based on the field operator algebra and conformal
representation theory; the other is canonical quantization
based on an action.
In the first approach, it is consistent to consider
one chiral sector of the usual Klein-Gordon field, and
to construct a set of primary fields which is closed under operator
product expansion \gins.
There are many consistent sets of fields with both integer and
fractional conformal dimensions.

As it will become clear in the following, we need an action
formulation in order to fix the field content for fractional Hall states.
Previous canonical approaches turn out to be insufficient for
our purposes. The Siegel action, quantized in ref.
\ref\imbi{C. Imbimbo and A. Schwimmer, {\it Phys. Lett.} {\bf B193}
(1987) 455.},
describes CFTs with integer dimensions only. A better candidate is
the first-order action of Floreanini and Jackiw \flore.
By quantizing it on a compact space, and by introducing and
additional coupling constant $\kappa$, we shall obtain sets of fields
with fractional dimensions appropriate for the fractional Hall effect.

Consider a real boson $\phi(x,t)$, living on a circle of length
$2\pi R$, and with dynamics governed by the action
\eqn\action{S\ =\ -{\kappa\over{4 \pi}}\ \int_{-\infty}^{+\infty}
dt\ \int_{0}^{2\pi R} dx\ \left({\partial}_{t} + v {\partial}_{x}
\right)\phi\ {\partial}_{x} \phi\ ,}
where $v$ is the ``light-velocity''. By rescaling the
space-time variables as $x\to R\theta$, $t\to Rt$, we can write
the corresponding Hamiltonian in the form
\eqn\chham{\eqalign{ H_R\ &=\ {v\over R} L_{0}\ , \cr
L_{0}\ &=\ {\kappa\over{4\pi}}\ \int_{0}^{2\pi} d\theta\
\left({\partial}_{\theta} \phi \right)^{2}\ ,\cr}}
where $L_0$ is again the dimensionless Hamiltonian of CFT.

Since the theory is defined on a compact space, special care is
needed in treating boundary conditions. We impose boundary
conditions
\eqn\solconf{\phi(2\pi ,t)-\phi (0,t)\ =\ -2\pi
\alpha_{0}\ ,}
parametrized by a time-independent constant $\alpha_{0}$.
As a consequence, the real boson $\phi$ has to be
considered as an {\it angular variable}, compactified on a
circle, and $\alpha_0$ becomes correspondingly quantized, as
we discuss below. By
considering variations that preserve \solconf , we find that
the action \action\ is minimized by fields that satisfy the
equation of motion
\eqn\eqmot{\left( {\partial}_t + v {\partial}_{\theta}
\right) {\partial}_{\theta} \phi\ =\ 0\ ,}
with boundary conditions
\eqn\bounc{\eqalign{ {\partial}_{\theta} \phi (2\pi,t)
\ &=\
{\partial}_{\theta} \phi (0,t)\ , \cr
\left( {\partial}_t + v {\partial}_{\theta}
\right) \phi (2\pi,t)\ &=\
\left( {\partial}_t + v {\partial}_{\theta}
\right) \phi (0,t)\ . \cr}}
The general solution of the equation of motion \eqmot\ can be
written as
\eqn\solemot{\phi(\theta,t)\ =\ f(\theta - v t)\ +\
g(t)\ .}
Here, $g(t)$ represents the gauge degree of freedom
corresponding to the invariance of the action \action\ and
the Hamiltonian \chham\ under $\phi\to\phi + \lambda(t)$.
Since it does not represent a physical degree of freedom, we discard
it by imposing the gauge condition
\eqn\gauco{\left( {\partial}_t + v {\partial}_{\theta}
\right) \phi\ =\ 0\ .}
In this gauge,
the general solution of \eqmot\ compatible with the boundary
conditions \solconf\ and \bounc ,
is given by
\eqn\gensol{\eqalign{
\phi(\theta-vt)\ &=\ \phi_{0} -\alpha_{0}
(\theta - vt)\ +\ i\sum_{n \ne 0}\ {{\alpha_{n}}\over n}
\ {\rm e}^{in(\theta - vt)}\ , \cr
{\alpha}^{*}_{n}\ &=\ {\alpha}_{-n}\ .\cr}}

Upon quantization, the coefficients $\phi_0$, $\alpha_0$ and
$\alpha_{n}$ of this expansion become operators acting on a
bosonic Fock space, whose vacuum $|\Omega\rangle$ is defined by
\eqn\bosvac{\alpha_{n} |\Omega\rangle\ =\ 0\  ,n > 0\ .}
The commutation relations of these operators are inferred from the
corresponding commutation relations of the field $\phi$ as
follows. The action \action\ is of {\it first order in time
derivatives}, and, therefore, the canonical momentum is completely
specified in terms of the field $\phi$ and its space derivative.
Nonetheless, there is {\it no need} to resort to the Dirac
formalism to quantize the theory, as was recently emphasized by
Faddeev and Jackiw
\ref\jack{L. Faddeev and R. Jackiw, {\it Phys. Rev. Lett.} {\bf 60}
(1988) 1692.}
. The canonical commutation relations can be
simply read off from the symplectic structure evident in the
action \action :
\eqn\eqtcom{ [ \phi(\theta , t), \phi({\theta}',t) ]
=\ i{\pi\over{\kappa}}\ \varepsilon(\theta - {\theta}')\ .}
These imply the commutation relations
\eqn\modcom{\eqalign{
[ \alpha_0 , \phi_0 ] &=\ {1\over{i\kappa}}\ ,\cr
[ \alpha_n , \alpha_m ] &=\ {n\over{\kappa}}\ \delta_{n+m,0}\ ,
\cr}}
for the coefficients of the expansion \gensol . In terms of these
operators, the Hamiltonian \chham\ takes the form
\eqn\hammod{L_0\ =\ {\kappa\over 2}{\alpha}^{2}_0\ +\
\kappa \sum_{n=1}^{\infty}\ \alpha_{-n}\alpha_{n}\ .}
Note that this expression has already been {\it normal ordered}, by
writing all the Fock space annihilators to the right. In analogy
with \mode , we also define the charges $L_n$, for $n \ne 0$, by
taking moments of the Hamiltonian density \chham\ :
\eqn\viraso{ L_{n}\ \equiv\ {\kappa\over 4\pi}\
\int_{0}^{2\pi} d\theta\ \left(\partial_\theta \phi\right)^2 \
e^{-in(\theta -vt)}\
=\ {\kappa\over 2}\ \sum_{l=-\infty}^{+\infty}\ {\alpha}_{n-l}
{\alpha}_{l}\ .}
By using the commutation relations \modcom ,
we finally obtain the chiral algebra of the bosonic theory:
\eqn\chalbo{\eqalign{[{\alpha}_{n},{\alpha}_{m}] &=\ {n\over \kappa}\
\delta_{n+m,0}\ , \cr
[L_{n},{\alpha}_{m}] &=\ -m\ \alpha_{n+m}\ , \cr
[L_{n},L_{m}] &=\ (n-m)\ L_{n+m}\ +\ {1\over 12}\ (n^{3} - n)\
\delta_{n+m,0}\ . \cr}}
This shows that the central charge is $c=1$.

In the following, we show that consistency of quantization
forces $\kappa$ to be {\it rational}.
Let us write $\kappa = p/q$, with $p$ and $q$ coprimes integers,
and choose $p>q$ without loss of generality.
{}From the boundary condition \solconf, $\phi (\theta)$ is a
topologically non-trivial mapping of the circle to itself.
We then choose the compactification radius of $\phi$
as $2\pi / p$,
\eqn\compac{\phi (\theta,t)\ \equiv\ \phi(\theta,t)\
+\ {2\pi \over{p}}\ ,}
which leads to the quantization condition
\eqn\topwind{\alpha_0\ =\ {n\over p}\ ,\ n \in {\bf Z}\ .}
Since $\kappa \alpha_0$ is the canonical momentum conjugate to
$\phi_0$, it follows that $\phi_0$ is an angular variable of range
$[0,2\pi q]$. This is also consistent with $\phi(\theta,t)$
being an angular variable of range $[0,{2\pi\over p}]$, due to
the rationality of $\kappa$.

We are now ready to discuss the operator content of the bosonic theory.
We first identify $\alpha_0$  with the total electromagnetic
charge (this identification will be justified in detail below).
Then we compute the grand-canonical partition
function
\eqn\gcbos{{\cal Z}(q,w)\ =\ \Tr\ q^{L_0 -{1\over 24}}\ w^{\alpha_0}
\ ,}
with $q$ and $w$ defined as in \grandc . Standard statistical
mechanics techniques give
\eqn\partfn{ {\cal Z}(q,w)\ =\ {1\over{\eta(q)}}
\ \sum_{n=-\infty}^{\infty}
q^{{1\over 2}{n^{2} \over{pq}}}\ w^{n\over{p}}\ ,}
where $\eta(q)$ is the Dedekind function \dedek\ and we have
used the quantization condition \topwind . For
$pq \ne 2 N^2$, the conformal dimensions\footnote{*}{For
$pq = 2 N^2$ we face the problem of degenerate representations of
the Virasoro algebra, and more care is required in extracting the
conformal dimensions of the primary fields \gins.}
and charges of the primary fields are thus
\eqn\haqbo{\eqalign{ h_{n}\ &=\ {n^{2} \over{2pq}}\ ,\cr
Q_{n}\ &=\ {n\over p}\ . \cr}}
The primary fields are the Fubini-Veneziano vertex operators
\eqn\vertex{V_{\beta}(\bar\eta) = \
:e^{i\beta \phi(\bar\eta)} : \ ,}
where $\bar\eta$ is the conformal variable in eq.\mapping.
These satisfy
\eqn\vercom{\eqalign{
[L_0 , V_{\beta} ]\ &=\left(
{\bar\eta}{\partial\over \partial\bar\eta }
\ +\ {{\beta}^{2} \over{2\kappa}} \right) V_{\beta}\ ,\cr
[ \alpha_0 , V_\beta ]\ &=\ {\beta\over\kappa}\ V_{\beta}\ ,\cr}}
confirming that $V_{\beta}$ is a primary field of conformal
dimension $h_{\beta} = {\beta}^{2} /2\kappa$ and charge $Q_{\beta}=
\beta / \kappa$. To obtain the primary fields of our theory, we have,
therefore, to restrict $\beta$ to the values
$\beta_n = n/q$. For later convenience, we
recall the operator product expansion of vertex operators:
\eqn\opeve{ V_{\beta_1}({\bar\eta}_1)\ V_{\beta_2}({\bar\eta}_2)\ =\
\left( {\bar\eta}_1 -{\bar\eta}_2
      \right)^{\beta_1 \beta_2 / \kappa}\
:V_{\beta_1} ({\bar\eta}_1) V_{\beta_2} ({\bar\eta}_2) :\ .}

We now discuss the special case $\kappa =1$. For $\kappa = 1$,
the operator content \haqbo\ of the bosonic theory coincides
with the operator content \haqsl\ of the fermionic theory
which describes edge excitations.
In particular, the fermion fields
are expressed by the bosonic vertex operators $V_{\pm 1}$,
since both are the primary fields of conformal dimension $1/2$
and charge $\pm 1$ in the respective representations:
$F=V_{-1}\ , \ F^{\dag} = V_{+1}$.
By using \opeve, one actually verifies the Minkowskian
anticommutators
\eqn\verequ{\eqalign{\{ V_{\pm 1}(\theta_1 - vt),
V_{\pm 1}(\theta_2 - vt) \}\ &=\ 0\ ,\cr
\{ V_{-1}(\theta_1 - vt),
V_{+1}(\theta_2 - vt) \}\ &=\ 2\pi\ \delta (\theta_1 -
\theta_2 )\ .\cr}}
The charge density on the cylinder
$\rho_R = :F^{\dag}_R F_R :$ can be accordingly expressed
in terms of the bosonic field. Normal ordering can be
implemented by subtracting the short-distance divergent part in
eq.\opeve, thereby obtaining
\eqn\rhobos{\rho_R =\ -{1\over 2\pi} {\partial}_x \phi \ ,}
which, in turn, implies the mapping ${\rho_n}={\alpha_n}\ $.
In particular, the total charge $\rho_0$ is represented in the
bosonic theory by $\alpha_0$, which justifies the previous
identification.

Having identified ${-{1\over 2\pi}}\partial_x \phi$ with the fermionic
charge density $\rho_R$ on the cylinder it is easy to couple
the chiral boson to $U(1)$ gauge fields.
This is accomplished by adding to the action \action\ the additional
piece
\eqn\deltas{\Delta S={e\over 2\pi}\
\int_{-\infty}^{+\infty}\ dt\ \int_0^{2\pi R}\ dx\
\left(A_t\partial_x \phi \ - \ A_x\partial_t \phi\right) \ \ .}
This coupling is Lorentz and gauge invariant (note that only
$A_{\mu}$'s are changed by gauge transformations).
Since the fermion current density is also $\rho_R$, due to the chiral
nature of the theory, eq.\deltas\ reduces to the standard coupling
$-A_{\mu} J^{\mu}$ in the physical gauge \gauco.
The equation of motion is changed to
\eqn\moto{\left( \partial_t +v\partial_x \right)
\left( {-{1\over 2\pi}}\partial_x \phi \right) \ = \
{e\over 2\pi\kappa} E \ ,}
with $E= \partial_t A_x - \partial_x A_t$ the electric field.
For $\kappa=1$ it reduces exactly to the chiral anomaly \anom\
of the fermionic representation.

We thus conclude that the $c=1$ CFT describing the edge dynamics
of one completely filled Landau level admits a bosonic representation.
This can be easily generalized by varying the rational parameter
$\kappa$.
\bigskip
\vfill\eject
\newsec{Fractional quantum Hall effect}
\bigskip
In this section, we address the physics of a fractionally filled
Landau level. Our approach is as follows.
We suppose an incompressible ground state is formed
at a rational value $\nu < 1$. Specifically, we consider the Laughlin
states at $\nu=1/m$, $m$ odd integer.
In these cases there exist short-range interactions such that
these states are the exact incompressible ground states \qhe.
Next, we assume that the universal long-range properties of these
states can be described by a $c=1$ CFT of interacting fermions,
or equivalently, by the chiral boson theory of the previous
section.
Presently, we are not able to derive directly this CFT by applying
the limit on the boundary of section $3$. Nevertheless, some
symmetry arguments support our assumption.

First, we can argue on general grounds that incompressibility in
the bulk should amount to conformal invariance on the boundary,
i.e. the ground state satisfies the highest-weight conditions
\vacpr\ $(\mu=1/2)$.
Indeed, in the case $\nu=1$ these conditions are equivalent \ctznext\
to the area-preserving ($W_\infty$-symmetry) conditions of the
bulk theory studied in ref.\ctz. There we showed that the $\nu=1/m$
Laughlin ground states also possess the same symmetry; therefore we
can expect again conformal invariance on the boundary.
Secondly, we expect that the central charge is again $c=1$.
Qualitatively, we know that the neutral excitations at $\nu=1/m$
with short-range potentials are the same as in the $\nu=1$ theory.
Following Laughlin, we only expect a modification of the charged
excitations, which should have Abelian statistics.

Actually, Laughlin's results can be derived on the single assumption of
central charge $c=1$. To this end we consider the bosonic
representation of section 5 and we identify the operator content
for a given $m$ as follows.
We first notice that the Hall current for
a state with filling fraction $\nu$ is
\eqn\hallcur{J^{i}\ =\ \nu{e\over{2\pi}}\ \epsilon^{ij} E^{j}\ .}
Given our previous identification of the Hall current with the
chiral anomaly of the CFT on the cylinder, we recognize
that the factor $\nu$ has to appear also in the chiral anomaly
equation
\eqn\anomn{(\partial_t\ +\ v\partial_x) \rho_R\ =\
\nu {e\over{2\pi}}\ E\ .}
Comparing this with the bosonic representation \moto\ of this
equation, we conclude that the effective CFT for the description
of an incompressible state with filling fraction $\nu$ is given
by \action\ with
\eqn\kapnu{\kappa\ =\ {1\over \nu}\ = \ m \  .}
We now explore the consequences of this identification.

First of all, we note that the {\it Casimir energy} $E_0 =
- v/24R$, and the corresponding quantum pressure $P= -v/48\pi
R^{3}$ of the electron liquid depend on the filling fraction
$\nu$ only through the velocities $v$. An independent measurement
of $P$ and $v$ for various filling fractions $\left( \nu= 1\ ,
{1\over 3}\ ,{1\over 5}\ \right)$ would thus provide an experimental
confirmation of the generic relationship between incompressibility
and conformal invariance for these states.

Given the identification \kapnu , we can recognize
the excitations about the incompressible state of filling
fraction $\nu=1/m$ from the partition function \partfn\ for
$p=m,q=1$.
These are neutral, gapless edge excitations (boundary phonons)
and quasi-particle and quasi-hole excitations with charges
$Q_{k} =\pm k/m$, and conformal dimensions $h_k = k^2/2m$,
with $k$ integer.
As in eq.\qhqp, they yield the energy spectra
\eqn\boupho{ \Delta E_{n,k}\ =\ \gamma (k) \ + \
{v \over R}{k^2\over 2m}\ +\ {v n\over R}\ .}
The first term is the non-universal gap for charged excitations
which can be again included by deforming the CFT as discussed in
section 3.

There are $2m$ basic charged excitations
associated to the primary fields $V_{\pm 1},\dots,V_{\pm m}$.
Repeating the arguments of section 4, we conclude that the
excitations about the incompressible ground state of filling
$\nu=1/m$ carry {\it fractional charge}
\eqn\frch{Q_k\ =\ \pm{k\over m}\ ,}
and {\it fractional statistics}
\eqn\frte{\theta_k\ =\ {k^{2} \over m}\quad{\rm mod}\ 2\ .}
Since $m =$ odd integer, there is
always the excitation $V_m$ with the quantum numbers $Q=1$ and
$\theta=1$ of an electron. These results are in perfect
agreement with Laughlin's theory, although they have been
derived from a completely different point of view.
Moreover, as already emphasized, conformal invariance leads to
predictions on finite size effects, which are beyond the scope
of the ``plasma analogy'', the main computational tool in Laughlin's
theory. These predictions are the universal Casimir energy
and the $1/R$-corrections to
the gaps for quasi-particle and quasi-hole excitations of
eq.\boupho.
Note that all these quantities have the same $(v/R)$-dependence.
Therefore, the ratios ($1/R$-parts)
\eqn\raten{ {\Delta E_{0,k}\over E_0 }\ =\ {12 k^2\over m}\ ,  \qquad
{\Delta E_{0,p}\over \Delta E_{0,q} }\ =\ { p^2\over q^2}\ ,  \qquad
k,p,q =1,\dots, m\ ,}
are universal quantities depending only on the filling fraction.
A measurement of these quantities, either by direct experiment or
by numerical simulation, would be a new and further confirmation
of the ideas presented in this paper.

We thus conclude by stressing the importance of finite-size
effects in the physics of the quantum Hall effect.
\bigskip
\noindent
{\bf Acknowledgments}
\bigskip
\noindent
We thank Sergio Fubini for inspiration and constant encouragement.
We also thank Luis Alvarez-Gaum\'e and J\"urg Fr\"ohlich for helpful
discussions. G. R. Z. thanks the World Laboratory for partial support.
G. V. D. gratefully acknowledges the support of the Mathematics Department
and the Center for Theoretical Physics at MIT, where some of this work
was done, and partial support through NSF grant 87-08447.
\vfill\eject
\appendix{A}
\bigskip
\noindent {\bf Direct derivation of long range order on the boundary}
\bigskip
In section $4$, we showed how conformal invariance implies algebraic
long-range order in the density matrix for points $z_1, z_2$ on the
boundary of the droplet.
Here we present the direct microscopic derivation of this result
for filling fraction $\nu=1$, thereby confirming the prediction of
the CFT.
Let us describe the steps of the derivation.

The density matrix is
\eqn\densm{\rho(z_1 ,z_2 )\ =\ \langle\Omega| {\Psi}^{\dag}
(z_1 ) {\Psi}(z_2 ) |\Omega\rangle\ ,\qquad
{\rm for}\ z_1 = R e^{i\theta_1} ,\ z_2 = R e^{i\theta_2}\ .}
We let $R\to\infty$ together with the number of particles,
$R^{2} = L + \mu$, taking for ease of computation
$eB=2$ ({\it i.e.}, $\ell=1$). The explicit
computation is based on the steepest-descent approximation of the
incomplete gamma function $\gamma(\alpha,x)$. One has
\eqn\frhoex{\eqalign{ \rho (z_1 ,z_2 )\ &=\ {1\over{\pi}}\
{\rm e}^{-{1\over 2} (|z_1|^{2} + |z_2 |^{2} )}\ \sum_{k=0}
^{L}\ {({\bar{z_1}}z_2 )^{k} \over{k!}}\ , \cr
&=\ {1\over{\pi}}\ {\rm e}^{R^{2}(e^{i\phi} - 1)}\ \left[
1\ -\ {R^{2(L+1)} \over{L!}}\ \int_{0}^{e^{i\phi}} dt\
{\rm e}^{L(-t+\log t)}\ {\rm e}^{-\mu t}\ \right]\ ,\cr}}
where $\phi=\theta_2 - \theta_1$ , and the integral in the
second term is taken along a ray from the origin to the point
$e^{i\phi}$ in the complex $t$-plane. We used an integral
representation of $\gamma (1+L,R^2 e^{i\phi})$, {\it i.e.},
of the confluent hypergeometric function \tables,
and made some changes of variables.
Following reference
\ref\asympt{A. Erd\'elyi, {\it ``Asymptotic Expansions''}, Dover,
New York (1956).},
the asymptotic expansion of the
integral for $L\to\infty$, is obtained by deforming the
integration contour along the steepest paths
\eqn\steep{Im\ h(t)\ =\ Im\ h(t_o)\ ,\qquad h(t)=-t+\log t ,\qquad
t=x+iy\ ,}
where $t_o$ is one of the two end-points of integration.

As a warming up exercise, let us first consider the case of the
density $\rho (z)$, {\it i.e.}, $z_1 = z_2$, $\phi =0$ and also
$\mu = 0$. The saddle point $t=1$ is one integration end
point, and the two steepest paths passing through it are the solutions of
\eqn\linesol{Im\ h(t)\ =\ -y\ + \arctan {y\over x}\ =\ 0\ ,\qquad
\to \qquad x={y\over{\tan y}}\ ,}
namely, the line $y=0$ and a non-trivial curve for $x<1$, ending
at $(x,y)=(0,\pm {\pi\over 2})$. The integration path in eq.
\frhoex\ coincides with the steepest path $y=0$, and it is not
deformed in this case. The saddle-point expansion of the integrand
$\exp h(t)$, including a factor $1/2$ for the half-path, gives
\eqn\integr{\int_{0}^{1} dt\ e^{L(-t+\log t )}\ \sim\ {1\over 2}
e^{-L}\ \left( {\sqrt{2\pi \over L}}\ +\ {4\over 3}{1\over L}\ +\
O \left( {1\over{L\sqrt{L}}} \right) \right) \ .}
Thus,
\eqn\resrho{\rho\left( |z|={\sqrt{L}} \right)\
=\ {1\over{\pi}}\ \left( {1\over 2}
\ -\ {2\over 3}{1\over{\sqrt{2\pi L}}}\ +\
O\left( {1\over{L}} \right) \right)\ .}
Actually, in the thermodynamic limit, the density of the
droplet becomes a representation of the step function
$\rho(z)={1\over \pi} \theta(L-|z|^{2} )$, whose mid-point
regularized value is $1/2$.

In the case of the density matrix, $\phi \ne 0$, the steepest
path going from  $t=0$ to $t=e^{i\phi}$, is the unique
solution to
\eqn\steepp{Im\ h(t)\ =\ Im\ h(e^{i\phi}) \qquad \to \qquad
x={y\over{\tan ( y+\phi -\sin \phi )}}\ .}
This path lies in the region $x<1$, and connects the three points
$(x,y)=(0,{\pi\over 2}-\phi+\sin \phi)$, $(\cos \phi ,\sin \phi )$
and $(0,0)$; so it is conveniently parametrized by $y$. One can
show that $Re\  h(t)$ is monotonically increasing in $y$, {\it i.e.},
along the path. Therefore, after deforming the integral along this
path, its asymptotic value is dominated by the integration region
around the end-point $t=e^{i\phi}$, where $Re \ h(t= x(y) +i y)$ can
be expanded to first order as
\eqn\result{\eqalign{
\int_{0}^{e^{i\phi}} & dt\ e^{L h(t)- \mu t} \cr
&\sim\ \exp\left( L(i\phi - e^{i\phi} ) -\mu e^{i\phi} \right)\
\int_{-\infty}^{\sin \phi} dy\
e^{2 L (y - \sin \phi )\tan {\phi \over 2}}\
\left( \left({dx\over{dy}}\right)_{y=\sin\phi} + i \right) \cr
&=\ \exp\left(L(i\phi - e^{i\phi} ) -\mu e^{i\phi} \right)\
{1\over L}{e^{i\phi} \over{1-e^{i\phi}}} .\cr}}
By collecting all factors, the asymptotic form of the density
matrix \densm\ for points on the boundary is
\eqn\correlb{\langle\Omega |{\Psi}^{\dag}(Re^{i\theta_1})
\Psi (Re^{i\theta_2})\ |\Omega\rangle\ \sim {1\over{\pi}}
{1\over\sqrt{2\pi} R}\ e^{i(R^{2}-\mu)(\theta_2 -\theta_1)}\
{e^{i\theta_2} \over{e^{i\theta_2} - e^{i\theta_1}}}\ ,
\quad (R\to\infty).}
Upon using $R^2 - \mu =L$, this is in agreement with the
prediction \twopom of the CFT.

\listrefs
\bigskip
\noindent{\bf Figure captions}
\bigskip
\bigskip
\noindent{\bf Fig. 1}

\noindent{ The structure of the filled first Landau level in the disk
geometry. All states up to angular momentum $L$ are occupied.}
\bigskip

\noindent{\bf Fig. 2}

\noindent{The expectation value of the charge density in units
$1/\ell^2$ for the first Landau level filled up to $L=50$ as a
function of $r/\ell$.}
\bigskip

\noindent{\bf Fig. 3}

\noindent{The expectation value of the transverse current in units
$1/(- 2m\ell^3)$ as a function of $r/\ell$ for the first Landau level
filled up to $L=50$.}
\bigskip

\noindent{\bf Fig. 4}

\noindent{Typical one-particle spectrum for an annulus geometry of
radii $R_1\simeq \ell\sqrt{L_1}$ and $R_2\simeq \ell\sqrt{L_2}$.
Full dots denote occupied levels.}
\bigskip

\noindent{\bf Fig. 5}

\noindent{Typical one-particle spectrum for the disk geometry of
radius $R\simeq \ell\sqrt{L}$. See eq.\betazero.}
\bigskip

\noindent{\bf Fig. 6}

\noindent{The one-body spectrum $\varepsilon_l$ in eq.\espect\
for $\Phi=10$.}
\bigskip

\noindent{\bf Fig. 7}

\noindent{The conformal map from the Minkowskian cylinder $(R\theta,t)$
to the conformal plane $\eta$, see eq.\mapping.}
\vfill
\end